\documentclass[10pt,twocolumn]{IEEEtran}

\usepackage{adjustbox}

\usepackage{diagbox}
\usepackage{enumitem}
\usepackage{amsmath,graphics,amssymb,epsfig,subfigure,color,cite}
\usepackage{array}
\usepackage{multirow}
\usepackage{enumerate}
\usepackage{algorithmic}
\usepackage{algorithm}
\usepackage{bm}
\usepackage{float}
\usepackage{makecell}
\usepackage{lipsum}
\usepackage{capt-of}

\floatname{algorithm}{Procedure}

\begin{document}
	
        \title{Joint Source-Channel Coding for Channel-Adaptive Digital Semantic Communications}
	 \author{Joohyuk Park, Yongjeong Oh, \IEEEmembership{Graduate Student Member,~IEEE}, Seonjung Kim, and Yo-Seb Jeon, \IEEEmembership{Member,~IEEE}
    \thanks{Joohyuk Park, Yongjeong Oh, Seonjung Kim, and Yo-Seb Jeon are with the Department of Electrical Engineering, POSTECH, Pohang, Gyeongbuk 37673, South Korea (e-mail: joohyuk.park@postech.ac.kr; yongjeongoh@postech.ac.kr; seonjung.kim@postech.ac.kr; yoseb.jeon@postech.ac.kr).}
    }
	\vspace{-2mm}
	
	\maketitle

	\vspace{-12mm}

 
    \begin{abstract} 

    In this paper, we propose a novel joint source-channel coding (JSCC) approach for channel-adaptive digital semantic communications. In semantic communication systems with digital modulation and demodulation, robust design of JSCC encoder and decoder becomes challenging not only due to the unpredictable dynamics of channel conditions but also due to diverse modulation orders. To address this challenge, we first develop a new demodulation method which assesses the uncertainty of the demodulation output to improve the robustness of the digital semantic communication system. We then devise a robust training strategy which enhances the robustness and flexibility of the JSCC encoder and decoder against diverse channel conditions and modulation orders. To this end, we model the relationship between the encoder's output and decoder's input using binary symmetric erasure channels and then sample the parameters of these channels from diverse distributions. We also develop a channel-adaptive modulation technique for an inference phase, in order to reduce the communication latency while maintaining task performance. In this technique, we adaptively determine modulation orders for the latent variables based on channel conditions. Using simulations, we demonstrate the superior performance of the proposed JSCC approach for image classification, reconstruction, and retrieval tasks compared to existing JSCC approaches. 
    \end{abstract}

    
	


        \section{Introduction}\label{Sec:Intro}
            Semantic communication \cite{SC_1, SC_2, SC_3} has garnered increasing attention, referring to the process of transmitting and receiving messages designed to convey meaning.
            In traditional communication, the focus of a transmitter is to transform the message into a bit sequence that can be accurately reconstructed at a receiver with minimal bit errors.
            In contrast, in semantic communication, the goal of the transmitter is to convey the meaning of the message, aiming to maximize the performance of a desired task at the receiver. 
            The primary advantage of semantic communication lies in its ability to enhance the task performance at the receiver, even in scenarios where perfect reconstruction of the bit sequence is not feasible using traditional communication systems.
            For example, in \cite{DeepJSCC, DeepJSCC-f, DeepJSCC-Multi_Task, DeepJSCC-MIMO}, it was reported that the semantic communication achieves a better task performance than the traditional communication that only focuses on delivering the bit sequence. 
            Motivated by this advantage, the semantic communication has been recognized as a crucial technology for enabling high-volume data-intensive and low-latency tasks such as VR/AR, video signal used in autonomous vehicle \cite{Application_2}, and a drone performing a specific mission \cite{Application_3}.

            There is a rich literature on the design of the semantic communication system that can deliver task-reliant information quickly and accurately over the wireless channels. 
            One notable approach involves employing a joint source-channel coding (JSCC) neural network, sometimes referred to as joint semantic channel coding. 
            In this approach, source and channel encoders/decoders are integrated into a unified neural-type encoder/decoder. Subsequently, the JSCC encoder and decoder are jointly trained by considering the impact of wireless channels such as additive white Gaussian noise (AWGN) and Rayleigh fading channels.  
            The design of the JSCC encoder/decoder was studied for various applications such as image transmission \cite{image_trans_1, image_trans_2, image_trans_3}, text transmission \cite{DeepJSCC-T,DeepSC, ReAllo-T}, speech transmission \cite{DeepSC-S}, and video transmission \cite{Video, Video_2}. 
            Through these studies, the potential of the JSCC approach to enhance task performance was demonstrated when compared to traditional separate source and channel coding approaches. 
           Unfortunately, these studies rely on the end-to-end training of the JSCC encoder and decoder for specific training environments. Consequently, their effectiveness may be compromised when operating in diverse communication environments that significantly differ from the training environments.

            
            To address this challenge, several studies have explored the concept of a channel-adaptive JSCC approach, enhancing adaptability to diverse channel environments.   
            A channel-adaptive JSCC approach for wireless image transmission was studied in \cite{ADJSCC}, which utilizes attention modules to extract feature importance based on signal-to-noise ratio (SNR)  and loss.
            This idea was extended in \cite{CA-JSCC,DeepJSCC-OFDM} by incorporating orthogonal frequency division multiplexing (OFDM) waveforms. 
            Also, in \cite{JSCC-ARC}, the attention mask was transformed into a binary mask for adaptive rate control. 
            The common idea behind these methods is to utilize channel information, such as SNR, as an additional input to control the JSCC encoder and decoder. This concept is realized by incorporating dedicated modules like attention modules or SNR-adaptive modules. However, this approach escalates both the network size and training complexity. 
            In \cite{edge}, channel-adaptive training was conducted by adjusting the number of active outputs of the encoder according to channel conditions. 
            Unfortunately, the aforementioned techniques primarily focus on analog communication systems, where encoder outputs, either real-valued or complex-valued, are transmitted using analog modulation. Consequently, these methods lack compatibility with modern digital communication systems. Moreover, implementing these techniques introduces various challenges, including issues related to the cost, size, and flexibility of RF hardware components.

            A few studies have attempted to integrate the JSCC approach into a digital semantic communication system. In \cite{revision_ref_1}, \cite{revision_ref_2}, a digital semantic communication system was employed where a bit sequence is generated through quantization and transmitted without digital modulation. In \cite{JCM}, the real-valued output of the JSCC encoder was mapped to binary phase shift keying symbols, while in \cite{DeepJSCC-Q}, \cite{revision_ref_3} a quantizer was used to convert the encoder's output into conventional quadrature amplitude modulation (QAM) symbols.
            Additionally, an adaptive masking strategy in \cite{revision_ref_4} was employed for robust operation of digital semantic communications in the presence of semantic noise.
            In \cite{image_trans_4}, the idea of robust information bottleneck was introduced to enable robust model training across various SNR levels. The common limitation of these techniques is the use of a fixed modulation scheme during the training of the JSCC encoder and decoder. Consequently, the trained encoder and decoder have no compatibility with other modulation schemes, except the one considered during the training process. Although an adaptive modulation technique in \cite{BER_adaptive} can be adopted to provide compatibility with multiple modulation schemes, this technique did not consider an end-to-end JSCC training  approach to account for the effects of fading channels and noise. To the best of the authors' knowledge, no previous studies have explored a JSCC approach for channel-adaptive digital semantic communications, despite its practical appeal for enabling the early adaptation of semantic communications.

    \begin{figure*}[t]
        \centering 
            {\epsfig{file=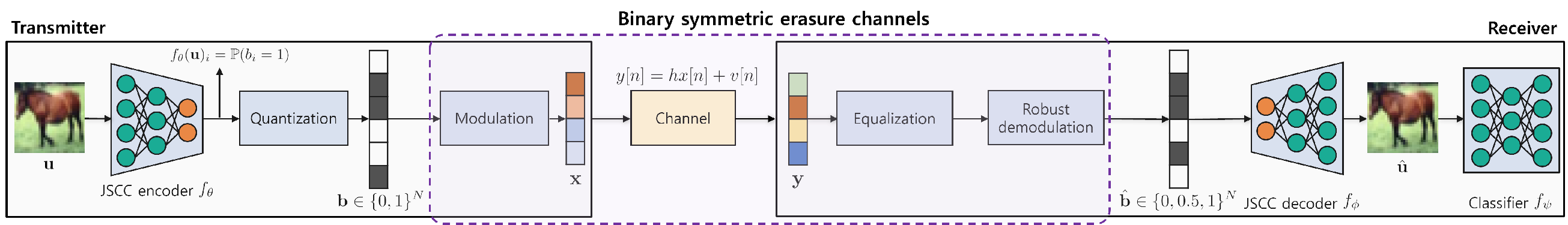, width=18cm}} 
        \caption{Illustration of the digital semantic communication system with the BSEC modeling considered in our work.}
        \label{fig:System}
    \end{figure*}  

    To bridge this research gap, this paper proposes a novel JSCC approach for channel-adaptive digital semantic communications, providing robustness and flexibility against diverse channel conditions and modulation schemes.
    The proposed approach comprises three novel components: (i) a robust demodulation method, (ii) a robust training strategy, and (iii) a channel-adaptive modulation technique. In the robust demodulation method, we assess the uncertainty of the demodulation output and assign an intermediate value instead of conventional binary outputs when uncertainty arises. Utilizing this method, we enhance the robustness of digital semantic communication systems against fading channels and noise.  In the robust training strategy, we model the relationship between the encoder's output and the decoder's input using binary symmetric erasure channels (BSECs) and then sample the parameters of these models from diverse distributions. By doing so, we not only facilitate end-to-end training of the JSCC encoder and decoder but also enhance their robustness and flexibility against diverse channel conditions and modulation orders. In the channel-adaptive modulation technique, we adaptively determine modulation orders of the latent variables according to channel conditions, thus reducing the communication latency for transmission while maintaining task performance.
    Using simulations, we demonstrate the superiority of the proposed JSCC approach for image classification, reconstruction, and retrieval tasks compared to existing JSCC approaches.
    The major contributions of our paper are summarized below.
    \begin{itemize}
        \item We develop a new demodulation method for improving the robustness of the digital semantic communication system. In this method, we introduce a criterion to assess the uncertainty of the demodulation output based on a log-likelihood ratio (LLR). We then assign an intermediate output $0.5$, instead of conventional binary outputs $0$ or $1$, when uncertainty arises. To reduce the computational complexity of the demodulation method, we also devise closed-form decision boundaries to check the uncertainty criterion. Through the design of the new demodulation method for semantic communications, we address the challenges posed by conventional hard-output demodulation considered in the literature (e.g., \cite{NECST}), which has limited expressive power in the latent space and is vulnerable to bit-flip errors in low-SNR regimes.

        \item We present a robust end-to-end training strategy for the JSCC encoder and decoder when employing our demodulation method. In this strategy, we employ BSECs to model the stochastic interaction between the encoder's output and the decoder's input. We then develop a sampling strategy which introduces variations in the bit-flip probabilities of the BSECs by sampling them from different stochastic distributions. Through this stochastic model with parameter sampling, our strategy effectively enhances the robustness and flexibility of the JSCC encoder and decoder against diverse channel conditions and modulation orders, in comparison to conventional environment-specific training strategies.


        \item We devise a channel-adaptive modulation technique for an inference phase, in order to reduce the communication latency while maintaining task performance. To this end, we characterize the bit-error and correct-decision probabilities of the QAM symbols as a function of the SNR and modulation order. Based on this characterization, we determine the best modulation order that can minimize the communication latency while ensuring that the bit-error probability of the QAM signal is below the bit-flip probability set by our training strategy. 

        \item Using simulations, we demonstrate the superiority of the proposed JSCC approach over the existing JSCC approaches for image classification, reconstruction and retrieval tasks using the MNIST \cite{MNIST}, Fashion-MNIST \cite{Fashion-MNIST}, CIFAR-10 and CIFAR-100 \cite{CIFAR-10} datasets. Our results show that the proposed approach outperforms the existing approaches in terms of the classification, reconstruction, and retrieval performances. Using simulations, we also validate the effectiveness of our demodulation method, training strategy, and channel-adaptive modulation technique. 
    \end{itemize}
    


    \section{System Model}\label{Sec:Model}
    In this work, we consider a digital semantic communication system for a dedicated machine learning task at a receiver. An example of the considered system for an image classification task is illustrated in Fig. \ref{fig:System}. 
    
    At the transmitter, a JSCC encoder configured with a deep neural network (DNN) is employed to transform an input image into a bit sequence. 
    Let $\bf{u}$ be an input data (e.g., an image) which is assumed to be independent and identically distributed (IID) over a source distribution $p_{\rm in}(\bf{u})$. The operation of the JSCC encoder is denoted by a function $f_{\theta}(\bf u)$ parameterized by the weights $\theta$. Suppose that the sigmoid function is employed as an activation function of the output layer. Then each output of the JSCC encoder can be interpreted as the probability
    of the modulated bit being 1, and the sampling from this distribution results in the generation of the bit sequence, ${\bf b} =[b_1,\cdots,b_N]^{\sf T}\in\{0,1\}^{N}$ where $N$ is the length of the bit sequence. The entries of the bit sequence $\bf b$ will be treated as {\em binary} latent variables. 
    After this, digital modulation is applied to transform the bit sequence ${\bf b}$ into a symbol sequence ${\bf x}$.
    The $n$-th entry of the symbol sequence ${\bf x}$ is denoted as $x[n]$, and this symbol is transmitted at time slot $n$.
    We assume that each symbol is modulated using $2^M$-QAM, i.e., $x[n] \in \mathcal{C}_M$, where $\mathcal{C}_M$ is a constellation set of $2^M$-QAM.

    The wireless channel of the system is modeled as quasi-static fading channels (also known as block fading channels) \cite{Block_fading}, in which channel coefficients remain constant within a channel coherence time.
    Under the quasi-static fading channel model, the baseband received signal at time slot $n$ is expressed as
    \begin{align}\label{eq:received_signal}
        y[n] = hx[n] + v[n],
    \end{align}
    where $h \in \mathbb{C}$ is a complex-valued channel coefficient, and $v[n] \in \mathbb{C}$ is an AWGN distributed as $\mathcal{CN}(0,\sigma^2)$.
    Suppose that the channel coefficient $h$ is perfectly estimated at the receiver via pilot-assisted channel estimation within every channel coherence time.
    By utilizing the knowledge of $h$, the channel equalization is executed for the received signal in \eqref{eq:received_signal}, which yields the equalized signal at time slot $n$ given by 
    \begin{align}\label{eq:equalized_signal}
        \tilde{y}[n] = \frac{h^*}{|h|^2}y[n] = x[n] +\tilde{v}[n],
    \end{align}
    where $\tilde{v}[n]\sim \mathcal{CN}(0,1/{\sf SNR})$ and ${\sf SNR} \triangleq |h|^2/\sigma^2$ is the instantaneous SNR of the system. 
    Then digital demodulation is executed to reconstruct the transmitted bit sequence from the equalized signals. 
    Details of a demodulation method adopted in our work will be introduced in Sec.~\ref{Sec:Demod}.
    By applying the aforementioned demodulation process, the transmitted bit sequence ${\bf b}$ is reconstructed at the receiver, which is denoted by $\hat{\bf b} = [\hat{b}_1,\cdots,\hat{b}_N]^{\sf T}$.
    After reconstructing the bit sequence, the JSCC decoder configured with a DNN is applied to reconstruct the input data ${\bf u}$, denoted by $\hat{\bf u}$. 
    The operation of the JSCC decoder is denoted by a function $\hat{\bf u} = f_{\phi}(\hat{\bf b})$ parameterized by the weights ${\phi}$.
    Finally, the dedicated machine learning task is performed by a task neural network $f_{\psi}$ (e.g., classifier), parameterized by the weights ${\psi}$, based on the reconstructed data $\hat{\bf u}$.

    \section{Robust Demodulation Method for Digital Semantic Communications}\label{Sec:Demod}
    In this section, we design a special type of demodulation, referred to as \textit{robust demodulation}, for improving the robustness of the digital semantic communication system described in Sec. II.
    
    \subsection{Design Principle}\label{Sec:Demod1}
    In traditional digital communication systems, two types of demodulation are typically considered: (i) soft-output demodulation, which  yields LLR values, and (ii) hard-output demodulation, which  generates binary outputs. Unfortunately, when applied to JSCC-based digital semantic communication systems, both demodulation methods encounter their own limitations as described below.
    {\begin{itemize}
    \item {\bf Limitations of soft-output demodulation:} The LLR values obtained from the soft-output demodulation have an infinite number of possibilities. Consequently, the statistical behavior of these values may involve an infinite number of parameters, with distributions influenced by both channel conditions and modulation orders. Therefore, integrating soft-output demodulation into the training of the JSCC encoder/decoder poses a considerable challenge in enabling the encoder/decoder to learn the diverse behavior of the LLR values formed under varying channel conditions and modulation orders. 

    \item {\bf Limitations of hard-output demodulation:} 
    The outputs of the hard-output demodulation are limited to binary values, significantly restricting the expressive power in the latent space. Furthermore, hard-output demodulation is susceptible to bit-flip errors, especially in low-SNR regimes, which may significantly alter the desired meaning of transmitted data.
\end{itemize}

    To address the limitations of the conventional demodulation methods, we devise a robust demodulation method that produces ternary outputs. This method introduces an intermediate value, denoted as $0.5$\footnote{Our intuition behind the choice of $0.5$ is that biasing the intermediate value towards a particular binary value (i.e., $0$ or $1$) might create difficulty for the JSCC decoder in distinguishing between the intermediate value and the corresponding binary value. This potentially leads to performance degradation in the overall communication process.}, in addition to the conventional binary values of $0$ and $1$. Specifically, our method assigns this intermediate value when uncertainty arises regarding the transmitted binary latent variable. This strategy effectively mitigates frequent bit-flip errors in low-SNR regimes while simultaneously enhancing the expressiveness of the demodulation output compared to conventional hard-output demodulation. These intrinsic features of our demodulation method significantly augment the JSCC decoder's capability to perform dedicated machine learning tasks. The advantage of our robust demodulation method in facilitating robust training of the JSCC encoder/decoder will be discussed in Sec.~\ref{Sec:Training}.


    In our demodulation method, we measure the reliability level of the decision on each latent variable based on the magnitude of the LLR, in order to determine a criterion for assigning the intermediate value $0.5$. 
    Let $n_m$ be the $m$-th binary latent variable associated with the transmitted symbol $x[n]$ at time slot $n$. 
    Also, let $\mathcal{C}_{m,u}^{(M)}$ be a subset of $\mathcal{C}_M$ which consists of symbols whose $m$-th bit is given by $u$ after a symbol demapping, i.e.,
    \begin{align}\label{eq:X_demap}
        \mathcal{C}_{m,u}^{(M)} = \big\{ c \in \mathcal{C}_M  ~\big|~ u_m = u, {\sf Demap}_M({c})=[u_1,\cdots,u_M] \big\},
    \end{align}
    for $u\in\{0,1\}$, where ${\sf Demap}_M(\cdot):\mathcal{C}_M \rightarrow \{0,1\}^M$ is a symbol demapping function for $2^M$-QAM. 
    Then the LLR of the $m$-th binary latent variable $b_{n_m}$ is computed as
    \begin{align}\label{eq:APP1}
        \mathcal{L}_m(\tilde{y}[n]) &= \ln \bigg(\frac{\mathbb{P}(b_{n_m}=0|\tilde{y}[n])}{\mathbb{P}(b_{n_m}=1|\tilde{y}[n])}\bigg) \nonumber \\ 
        &= \ln \bigg(\frac{\sum_{c^\prime \in \mathcal{C}_{m,0}^{(M)}} \mathbb{P}({x}[n]={c}^\prime|\tilde{y}[n])}
        {\sum_{c \in \mathcal{C}_{m,1}^{(M)}} \mathbb{P}({x}[n]={c}|\tilde{y}[n])}\bigg) \nonumber \\
        &= \ln \bigg(\frac{\sum_{c^\prime \in \mathcal{C}_{m,0}^{(M)}} \mathbb{P}({x}[n]={c}^\prime, \tilde{y}[n])}
        {\sum_{c \in \mathcal{C}_{m,1}^{(M)}} \mathbb{P}({x}[n]={c}, \tilde{y}[n])}\bigg).
    \end{align}
    Unfortunately, prior knowledge about the distribution of bit outputs from the JSCC decoder may not be available at the receiver because it depends on the true distribution of both the encoder weights and the source data. To circumvent this challenge, we assume that the prior probability of each binary latent variable is uniform. Under this assumption, the LLR in \eqref{eq:APP1} is rewritten as
    \begin{align}\label{eq:APP2}
        \mathcal{L}_m(\tilde{y}[n]) &= \ln \Bigg(\frac{\sum_{c^\prime \in \mathcal{C}_{m,0}^{(M)}} p(\tilde{y}[n]|{x}[n]={c}^\prime)}
        {\sum_{c \in \mathcal{C}_{m,1}^{(M)}} p(\tilde{y}[n]|{x}[n]={c})}\Bigg)  \nonumber \\
        &\overset{(a)}{=} \ln \Bigg(\frac{\sum_{c^\prime \in \mathcal{C}_{m,0}^{(M)}} \exp(-{{\sf SNR}|(\tilde{y}[n]-c^\prime)|^2})}
        {\sum_{c \in\mathcal{C}_{m,1}^{(M)}} \exp(-{{\sf SNR}|(\tilde{y}[n]-c)|^2})}\Bigg),
    \end{align}
    where $(a)$ follows from \eqref{eq:equalized_signal}. 
    Note that if $\mathbb{P}(b_{n_m}=0|\tilde{y}[n]) \approx \mathbb{P}(b_{n_m}=1|\tilde{y}[n])$, the demodulation is uncertain about its decision on $b_{n_m}$ and the corresponding LLR will be close to zero.
    Therefore, we use the magnitude of the LLR $\mathcal{L}_m(\tilde{y}[n])$ as a measure of the reliability level when making a decision about the binary latent variable $b_{n_m}$ based on the observation $\tilde{y}[n]$. 
    In particular, our demodulation method assigns the intermediate value $0.5$ to the $m$-th latent variable $b_{n_m}$ whenever the corresponding LLR is close to $0$.
    Let $\rho_{n_m} > 0$ be a threshold applied to the LLR for assigning the intermediate value. 
    Then the output of our demodulation method can be expressed as ${\sf Demod}_M(\tilde{y}[n])= [\hat{b}_{n_1},\hat{b}_{n_2}, \cdots, \hat{b}_{n_M}]$, where
    \begin{align}\label{eq:demod}
        \hat{b}_{n_m} = \begin{cases}
        0, & \text{if}~\mathcal{L}_m(\tilde{y}[n]) > \rho_{n_m}, \\
        0.5, & \text{if}~|\mathcal{L}_m(\tilde{y}[n])| \leq \rho_{n_m}, \\
        1, & \text{if}~\mathcal{L}_m(\tilde{y}[n]) < -\rho_{n_m}.
        \end{cases}
    \end{align}

    \subsection{Low-Complexity Robust Demodulation Method}\label{Sec:Demod2}
    The robust demodulation method in \eqref{eq:demod} requires high computational complexity due to the need to measure distances to all symbols in $\mathcal{C}_M$. To alleviate the computational complexity, we design a low-complexity variation of the robust demodulation method. We start by approximating the LLR as
    \begin{align}\label{eq:APP3}
        \mathcal{L}_m(\tilde{y}[n])
        &\overset{(a)}{\approx} 
        \underset{c \in \mathcal{C}_{m,1}^{(M)}}{\rm min} \frac{|(\tilde{y}[n]-c)|^2}{1/{\sf SNR}}-\underset{c^\prime \in \mathcal{C}_{m,0}^{(M)}}{\rm min}\frac{|(\tilde{y}[n]-c^\prime)|^2}{1/{\sf SNR}},
    \end{align}
    where $(a)$ follows from a well-known log-sum-exp approximation. Thanks to the independence of the real and imaginary parts of the AWGN along with the symmetric property of the QAM, the demodulation method for the bits associated with the real and imaginary parts can be performed independently. Utilizing this fact, we define two LLR functions:
    \begin{align}\label{eq:LLR_I_Q}
        \mathcal{L}_{I,m}(\tilde{y}[n])
        =& \underset{c \in \mathcal{C}_{m,1}^{(M)}}{\rm min} \frac{(\mathfrak{Re}\left\{\tilde{y}[n]-c\right\})^2}{1/{\sf SNR}}  \nonumber \\
         &-\underset{c^\prime \in \mathcal{C}_{m,0}^{(M)}}{\rm min}\frac{(\mathfrak{Re}\left\{\tilde{y}[n]-c^\prime\right\})^2}{1/{\sf SNR}}, \nonumber \\
        \mathcal{L}_{Q,m}(\tilde{y}[n])
        =& \underset{c \in \mathcal{C}_{m,1}^{(M)}}{\rm min} \frac{(\mathfrak{Im}\left\{\tilde{y}[n]-c\right\})^2}{1/{\sf SNR}} \nonumber \\
        &-\underset{c^\prime \in \mathcal{C}_{m,0}^{(M)}}{\rm min}\frac{(\mathfrak{Im}\left\{\tilde{y}[n]-c^\prime\right\})^2}{1/{\sf SNR}},
    \end{align}
    which correspond to the bits associated with the real and imaginary parts, respectively.
    Depending on the value of $m$, a proper LLR function is chosen between $\mathcal{L}_{I,m}(\tilde{y}[n])$ and $\mathcal{L}_{Q,m}(\tilde{y}[n])$. In particular, our criterion for assigning the intermediate value in the second line of (6) is rewritten as 
    \begin{align}\label{eq:robust_detect_1}
        |\mathcal{L}_{I,m}(\tilde{y}[n])| \leq \rho_{n_m}~\text{or}~ |\mathcal{L}_{Q,m}(\tilde{y}[n])| \leq \rho_{n_m}.
    \end{align}
    

    To facilitate the low-complexity computation for checking the criterion in \eqref{eq:robust_detect_1}, we now introduce decision boundaries in an in-phase-quadrature (I-Q) constellation diagram for $2^M$-QAM.
    Let $a_{i}\in[0,1]$ be the first decision boundary, which is closest to the I-axis or the Q-axis, to check the above criterion for a given $\rho_i$, where the bit index $i$ satisfies $1+ \left \lfloor (i-1)/M \right \rfloor=n$. From \eqref{eq:LLR_I_Q} and \eqref{eq:robust_detect_1}, if the $i$-th bit has different values across $a_{i}$, then $\rho_i$ can be expressed as
    \begin{align}\label{eq:a_tau}
        \rho_{i}  &= {\sf SNR}\left\{\left(\frac{(a_{i}+1)d_M}{2}\right)^2-\left(\frac{(1-a_{i})d_M}{2}\right)^2\right\} \nonumber \\ 
        &= {\sf SNR}d_M^2a_{i} = \frac{6{\sf SNR}}{2^M-1}a_i,
    \end{align}
    where $d_M=\sqrt{{6}/{(2^M-1)}}$ is the minimum distance for the normalized $2^M$-QAM constellation set.
    Since adjacent decision boundaries maintain equidistant intervals, once $a_i$ is determined, we obtain the decision boundary values for $\forall j \in \{0,\pm1,\cdots,\pm{\sqrt{2^M}}/{2}\}$ as follows:
    \begin{align}\label{eq:equidistant}
        a_{i,j+1} &= \sqrt{\frac{2(2^M-1)}{3}}d_M + a_{i,j} = 2 + a_{i,j},
    \end{align}
    where $a_{i}=a_{i, 1}$.

    Let $D^U_{i,M,{0.5},j}$ and ${D}^L_{i,M,{0.5},j}$ be the sets of scaled upper and lower decision boundaries associated with two adjacent symbols having different $i$-th bit values. Similarly, let $D^U_{i,M,k,j}$ and $D^L_{i,M,k,j}$ be the sets of scaled upper and lower decision boundaries associated with two adjacent symbols having the same $i$-th bit value equal to $k \in \{0,1\}$.
    Utilizing these notations, the low-complexity robust demodulation for the real-part bits is represented as
    \begin{align}\label{eq:robust_demod_final}
        \hat{b}_{i} = k, ~ \text{if}~\mathfrak{Re}\left\{\tilde{y}[n]\right\} \in A_{i,M,k},
    \end{align}
    for $k \in \left\{0,0.5,1\right\}$, where
    \begin{align}\label{eq:robust_demod_final_2}
        A_{i,M,k} &= \bigcup_{j \in I_{i,M,k}}^{}\left\{u \in \mathbb{R}: D^L_{i,M,k,j} \leq u \leq D^U_{i,M,k,j} \right\}, \notag\\ 
        D^L_{i,M,k,j} &= \begin{cases}
        \frac{4(j-1)-a_{i,j}}{2}d_M, & \text{ if } k=0.5, \\ 
        \frac{a_{i,j-1}}{2}d_M, & \text{ if } k \in \left\{0,1\right\}, \notag 
        \end{cases}\\ 
        D^U_{i,M,k,j} &= \begin{cases}
        \frac{a_{i,j}}{2}d_M, & \text{ if } k=0.5, \\
        \frac{4j-a_{i,j+1}}{2}d_M, & \text{ if } k \in \left\{0,1\right\},
        \end{cases} 
    \end{align}
    where $I_{i,M,k} \in \mathbb{N}$ denotes the set of indices of decision boundaries related to the $i$-th bit being $k$ in the $2^M$-QAM.    

Fig.~\ref{fig:16QAM_detect} visualizes the decision boundaries associated with the second bit of a 16-QAM symbol (i.e., $i=2, M=4$). Note that in this case, we have $I_{2,4,0} = \left\{-1,3\right\}$, $I_{2,4,1} = \left\{1\right\}$ and $I_{2,4,0.5} = \left\{0,2\right\}$. 
In Fig.~\ref{fig:16QAM_detect}, we also visualize the regions $A_{2,4,0}, A_{2,4,1}$ and $A_{2,4,0.5}$ in \eqref{eq:robust_demod_final_2}, which are shaded in different colors. It should also be noted that the value of the second bit is determined as $0$ when $\mathfrak{Re}\left\{\tilde{y}[n]\right\} \leq D^U_{2,4,0,-1}$ or $\mathfrak{Re}\left\{\tilde{y}[n]\right\} \geq D^L_{2,4,0,3}$. Fig.~\ref{fig:16QAM_detect} clearly illustrates the decision rule of our demodulation method, which readily determines the demodulation output by comparing the value of the received signal with a given set of decision boundaries.
Due to the symmetric property of QAM, the low-complexity robust demodulation for the imaginary-part bits can be executed similarly. This involves replacing $\mathfrak{Re}\left\{\tilde{y}[n]\right\}$ with $\mathfrak{Im}\left\{\tilde{y}[n]\right\}$ and appropriately defining the decision boundary sets $\big\{D^U_{i,M,k,j}, D^L_{i,M,k,j}\big\}_{k \in \left\{0,0.5,1\right\}}$ for the imaginary-part bits.

    \vspace{3mm}

    \begin{figure}[t]
        \centering 
            {\epsfig{file=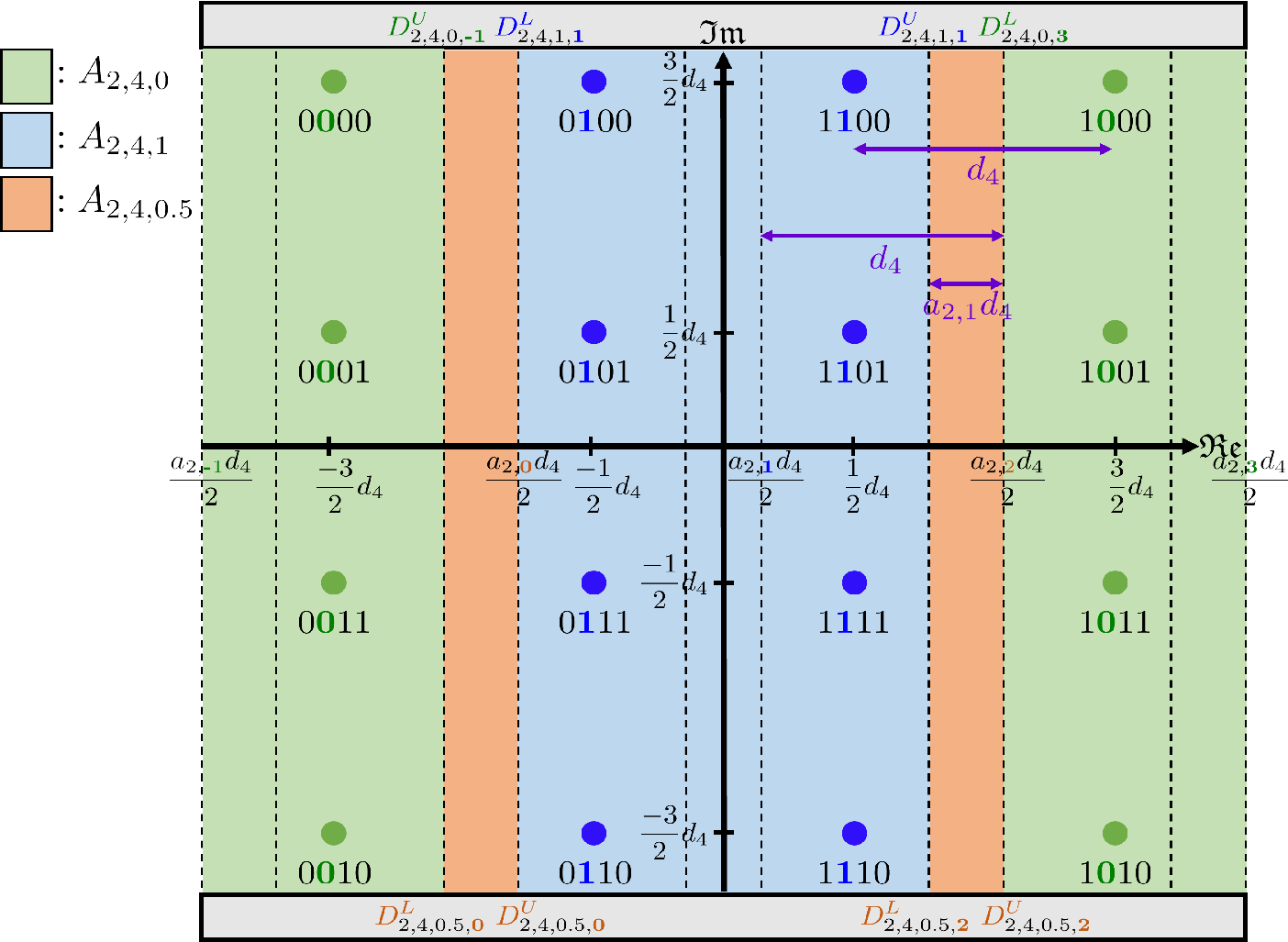, width=8.8cm}}
        \caption{Visualization of the decision boundaries associated with the second bit of a 16-QAM symbol.}
        \label{fig:16QAM_detect}
    \end{figure}

   {\bf Remark 1 (Demodulation with $K$-ary outputs):} Our demodulation method can be generalized to produce $K$-ary outputs for any $K>2$. As $K$ increases, the demodulation's expressive power grows, but controlling the outputs becomes more challenging due to the increased parameters needed to represent their statistical behavior. Moreover, the computational complexity of the demodulation process escalates with $K$. Therefore, we choose $K=3$ as a practical compromise, balancing model complexity and expressive power.

    \section{Robust Training Strategy of the Proposed JSCC Approach}\label{Sec:Training}
    In this section, we devise a robust training strategy to enhance the robustness and flexibility of the JSCC encoder and decoder against diverse channel conditions and modulation orders.
    

    \subsection{BSEC Modeling Approach}\label{Sec:BSEC}

    The fundamental idea of our robust training strategy is to harness a stochastic model to represent the combined effects of digital modulation, fading channel, equalization, and demodulation, instead of explicitly considering their effects individually.
    In the digital semantic communication system with our robust demodulation method in Sec. III-B, a binary latent variable $b_i \in \{0,1\}$ is stochastically transformed into a ternary variable $\hat{b}_i \in\{0,0.5,1\}$. 
    This stochastic transformation exactly follows a well-known BSEC model in which the intermediate value $0.5$ is treated as an erased value assigned when the symbol is erased. 
    Motivated by this fact, we harness $N$-parallel BSECs to model the relationship between the JSCC encoder's output $\bf{b}$ and the JSCC decoder's input $\hat{\bf{b}}$, as illustrated in Fig.~\ref{fig:System}. 
    In this model, the conditional distribution of the decoder input $\hat{b}_i$ for a given encoder output $b_i$ is represented as 
    \begin{align}\label{channel_model}
        p_{\mathrm{BSEC}}(\hat{b}_i|b_i;\tilde{\mu}_i, \tilde{d}_i) = 
        \begin{cases}
            \tilde{r}_i, & \text{if}~\hat{b}_i = b_i, \\
            \tilde{d}_i, & \text{if}~\hat{b}_i = 0.5, \\
            \tilde{\mu}_i, & \text{if}~\hat{b}_i \neq b_i, 
        \end{cases}
    \end{align}
    where $\tilde{r}_i+\tilde{d}_i+\tilde{\mu}_i = 1$, $\tilde{\mu}_i \in [0,1]$ represents the bit-flip probability, $\tilde{d}_i \in [0,1]$ represents the bit-erasure probability, and $\tilde{r}_i \in [0,1]$ represents the bit-correct probability for the $i$-th BSEC. 
    Note that if $\tilde{d}_i=0$, implying that there is no intermediate output $0.5$, the above BSEC reduces to a BSC which is one of the most widely adopted models in wireless communications. 
    Similarly, if $\tilde{\mu}_i=0$, implying that there is no bit-flip error, the above BSEC reduces to a binary erasure channel (BEC) which is another common model in wireless communications. Therefore, our BSEC model can be considered as a generalization of both the BSC and BEC models. By introducing not only the bit-flip probability but also the bit-erasure probability, BSEC offers the advantage of reducing the likelihood of bit-flip errors compared to the BSC, while allowing the consideration of the inevitable bit-flip errors that are ignored in the BEC.

    A key advantage of our BSEC modeling approach is that it facilitates the end-to-end training of the JSCC encoder and decoder without explicitly considering digital modulation, fading channel, channel equalization, and digital demodulation processes; rather, the combined effects of these processes are implicitly captured by the parallel BSECs with parameters $\{\tilde{\mu}_i,\tilde{d}_i,\tilde{r}_i\}_{i=1}^N$ during a training phase. 
    

    \subsection{Parameter Sampling Strategy}\label{Sec:Sample}
    When employing our BSEC modeling approach, it is crucial to determine a proper set of parameters $\{\tilde{\mu}_i,\tilde{d}_i,\tilde{r}_i\}_{i=1}^N$ that can reflect communication scenarios encountered during an inference phase.
    If we simply consider fixed parameters during a training phase, they can only represent a certain communication scenario with a particular channel condition and modulation level. For instance, if we only consider small bit-flip probabilities (i.e., $\tilde{\mu}_i \ll 1$) during the training phase, these probabilities may not align with communication scenarios where the SNR is low and/or the modulation order $M$ is high, since these scenarios may lead to high bit-error probabilities. 
    This motivates us to consider flexible determination of the parameters to ensure the robustness of the JSCC encoder and decoder against diverse SNRs and modulation orders that may occur during the inference phase.

    To promote flexibility in the parameter determination, our strategy is to introduce variations in the bit-flip probabilities $\{\tilde{\mu}_i\}_{i=1}^N$ by sampling them from different stochastic distributions. 
    In particular, when computing the loss for each training data, we sample the bit-flip probability $\tilde{\mu}_i$ of the $i$-th BSEC independently from the following uniform distribution\footnote{The uniform distribution is employed due to its simplicity; however, alternative distributions like the beta and exponential distributions, which sample values between 0 and 1, can also be applicable.}:
    \begin{align}\label{eq:sigmoid}
        \tilde{\mu}_i \overset{ \text{i.i.d} }{\sim} {\rm Uniform}[0, \alpha_i], 
    \end{align}
    where $\alpha_i \in [0, 0.5]$ is a target robustness level which represents the maximum bit-flip probability allowed for the $i$-th latent variable $b_i$. 
    By doing so, the latent variable $b_i$ can be trained to cover the bit-error probability up to the target robustness level $\alpha_i$.
    In our training strategy, we set different robustness levels across the latent variables, in order to allow flexibility in choosing different modulation orders, as will be discussed in Sec.~\ref{Sec:Modulation}.

    After sampling the bit-flip probabilities $\{\tilde{\mu}_i\}_{i=1}^N$, we determine the remaining parameters $\tilde{d}_i$ and $\tilde{r}_i$ that match with the sampled value of $\tilde{\mu}_i$.
    Note that three parameters $\tilde{d}_i$, $\tilde{r}_i$, and $\tilde{\mu}_i$ are entangled by the demodulation rule for a given SNR and modulation order. 
    However, both the SNR and modulation order are diverse during the inference phase, making it difficult to characterize the exact relationship among these parameters. 
    To circumvent this difficulty, during the training phase, we assume that 4-QAM (i.e., $M=2$) is chosen with $\rho_i = {\sf SNR}$. Under this assumption, the first decision boundary is given by $a_i=0.5$ for all $i\in\{1,\ldots,N\}$ by the relationship between $a_i$ and $\rho_i$ characterized in \eqref{eq:a_tau}. 
    Therefore, the bit-erasure probability $\tilde{d}_i$ associated with the sampled value of $\tilde{\mu}_i$ is determined by plugging $a_i=0.5$ and $\tilde{\sf SNR}_i(\tilde{\mu}_i)$ into the parameter characterization in \eqref{eq:flip_correct_final} and \eqref{eq:erasure_final}. As a result, the corresponding bit-erasure probability $\tilde{d}_i$ is given by
    \begin{align}
        \tilde{d}_i = {\sf Q}\left(\frac{1}{3}{\sf Q}^{-1}(\tilde{\mu}_i)  \right) - \tilde{\mu}_i,
    \end{align}
    where ${\sf Q}^{-1}(\cdot)$ is the inverse ${\sf Q}$-function.
    Similarly, the corresponding bit-correct probability $\tilde{r}_i$ is determined as $\tilde{r}_i = 1-\tilde{\mu}_i-\tilde{d}_i$. These parameter expressions are utilized during the training phase.

    A key advantage of the above sampling strategy is that the JSCC encoder and decoder can be trained with various realizations of $\{\tilde{\mu}_i,\tilde{d}_i,\tilde{r}_i\}_{i=1}^N$ during the training process.
    Therefore, our strategy effectively enhances the robustness of the JSCC encoder and decoder against diverse channel conditions that can be encountered during an inference phase. 
    This advantage will be numerically demonstrated in Sec.~\ref{Sec:Simul}. 

    \subsection{Training with BSEC Model}
    To train the JSCC encoder and decoder using the BSECs with diverse parameters, we modify the training strategy in \cite{NECST} which was originally developed for the BSCs with the same bit-flip probabilities.
    Following the strategy in \cite{NECST}, we aim at maximizing the mutual information between $\bf{u}$ and $\hat{\bf{b}}$.
    By increasing the dependency between $\bf{u}$ and $\hat{\bf{b}}$, the decoder gains the ability to effectively utilize $\hat{\bf{b}}$ for accurate prediction of $\bf{u}$ and therefore reduce the difference between $\bf{u}$ and $\bf{\hat{u}}$. 
    
    Let $\boldsymbol{\tilde{\mu}} = [\tilde{\mu}_1, \tilde{\mu}_2,\cdots,\tilde{\mu}_N]^{\sf T}$ and $\boldsymbol{\tilde{d}} = [\tilde{d}_1, \tilde{d}_2,\cdots,\tilde{d}_N]^{\sf T}$, where the $i$-th entry of $\boldsymbol{\tilde{\mu}}$ and $\boldsymbol{\tilde{d}}$ represents the bit-flip probability and bit-erasure probability determined by our sampling strategy in Sec.~\ref{Sec:Sample} for the $i$-th BSEC, respectively. 
    Then the conditional distribution of the parallel BSECs parameterized by $\boldsymbol{\tilde{\mu}}$ and $\boldsymbol{\tilde{d}}$ is expressed as 
    \begin{align}
        p_{\mathrm{channel}}(\hat{\bf{b}}|{\bf{b}}; \boldsymbol{\tilde{\mu}},\boldsymbol{\tilde{d}})
        &=\prod_{i=1}^{N}p_{\mathrm{BSEC}}(\hat{b}_i|b_i;\tilde{\mu}_i, \tilde{d}_i),
    \end{align}
    where each factor $p_{\mathrm{BSEC}}(\hat{b}_i|b_i;\tilde{\mu}_i, \tilde{d}_i)$ is characterized in \eqref{channel_model}.
    Meanwhile, by assuming that each bit $b_i$ follows an independent Bernoulli distribution with probability $f_{\theta}({\bf{u}})_i$, the conditional distribution of the JSCC encoder $p_{\theta}({\bf{b}}|{\bf{u}})$ follows
    \begin{align}
        p_{\theta}({\bf{b}}|{\bf{u}}) &= \prod_{i=1}^{N}(f_{\theta}({\bf{u}})_i)^{b_i}(1-f_{\theta}({\bf{u}})_i)^{1-b_i}.
    \end{align}
    Generating $b_i$ via sampling from the Bernoulli distribution can be regarded as stochastic 1-bit quantization for the $i$-th encoder output which is originally a real-valued constant.
    The distributions $p_{\theta}({\bf{b}}|{\bf{u}})$ and $p_{\mathrm{channel}}(\hat{\bf{b}}|{\bf{b}}; {\boldsymbol{{\tilde{\mu}}}, \boldsymbol{{\tilde{d}}}})$ imply that the channel from ${\bf{u}}$ to $\hat{\bf{b}}$ can be described as a noisy memoryless channel with the following conditional distribution \cite{Informax}:
    \begin{align}
        p_{\rm noisy}(\hat{\bf{b}}|{\bf{u}}; \boldsymbol{\tilde{\mu}}, \boldsymbol{\tilde{d}}, \theta) &= \sum_{{\bf{b}}\in \left\{0,1 \right\}^N}^{}p_{\theta}({\bf{b}}|{\bf{u}})p_{\mathrm{channel}}(\hat{\bf{b}}|{\bf{b}}; {\boldsymbol{{\tilde{\mu}}}, \boldsymbol{{\tilde{d}}}}) \nonumber\\
        &= \prod_{i=1}^{N} \underbrace{\sum_{b_i \in \left\{0,1 \right\}}^{}p_{\theta}(b_i|{\bf{u}})p_{\mathrm{BSEC}}(\hat{b}_i|b_i; \tilde{\mu}_i,\tilde{d}_i)}_{ = p_{\rm noisy}(\hat{b}_i|{\bf{u}}; \tilde{\mu}_i, \tilde{d}_i, \theta)}, \label{eq: noisy_channel}
    \end{align}
    where $p_{\rm noisy}(\hat{b}_i|{\bf{u}}; \tilde{\mu}_i, \tilde{d}_i, \theta)$ is further computed as 
    \begin{align} \label{eq: noisy_channel_2}
        &p_{\rm noisy}(\hat{b}_i|{\bf{u}}; \tilde{\mu}_i,\tilde{d}_i, \theta) \nonumber\\ 
        &= \sum_{b_i \in \left\{0,1 \right\}}^{}p_{\theta}(b_i|{\bf{u}})p_{\mathrm{BSEC}}(\hat{b}_i|b_i; \tilde{\mu}_i,\tilde{d}_i) \nonumber  \\
        &=
        \begin{cases}
            \tilde{\mu}_if_{\theta}({\bf{u}})_i +(1-\tilde{d}_i-\tilde{\mu}_i)(1-f_{\theta}({\bf{u}})_i),  &\text{ if } \hat{b}_i = 0, \\
            \tilde{d}_i,  &\text{ if } \hat{b}_i = 0.5, \\
            \tilde{\mu}_i(1-f_{\theta}({\bf{u}})_i)+(1-\tilde{d}_i-\tilde{\mu}_i)f_{\theta}({\bf{u}})_i,  &\text{ if } \hat{b}_i = 1.\\
        \end{cases}
    \end{align}
    During the training process, the decoder's input $\hat{\bf{b}}$ is sampled according to \eqref{eq: noisy_channel_2} for each input data ${\bf{u}}$. Note that by employing the sampling strategy in Sec.~\ref{Sec:Sample}, the parameters $\tilde{\mu}_i$ and $\tilde{d}_i$ are independently sampled across different input data.

    The mutual information $I(\bf{u}; \bf{\hat b})$  between $\bf{u}$ and $\hat{\bf{b}}$ is computed as
    \begin{align}
        &I(\bf{u}; \hat{\bf{b}}) =  \mathit{H}({\bf{u}})-\mathit{H}({\bf{u}}|\hat{\bf{b}}) \nonumber\\
        &= \mathit{H}({\bf{u}})+\mathbb{E}_{{\bf{u}} \sim p_{\rm in}({\bf{u}})} \big[\mathbb{E}_{\hat{\bf{b}} \sim p_{\rm noisy}(\hat{\bf{b}}|\bf{u}; {\boldsymbol{\tilde{\mu}}}, {\boldsymbol{\tilde{d}}}, \theta)}[\log p(\bf{u}|\hat{\bf b})] \big],
    \end{align}
    where $p(\bf{u}|\hat{\bf b})$ is the true posterior distribution. 
    Unfortunately, this distribution is often intractable, so we cannot directly train the JSCC decoder to follow the true posterior distribution. To circumvent this limitation, we use a variational approximation  $p_{\phi}({\bf u}|\hat{\bf b})$ \cite{JCM} by assuming that the JSCC decoder $f_{\phi}(\hat{\bf b})$ is a stochastic decoder whose output follows a Gaussian distribution with the mean $f_{\phi}(\hat{\bf b})$ and an isotropic covariance matrix (i.e., $p_{\phi}({\bf u}|\hat{\bf b}) = \mathcal{N}({\bf u}; f_{\phi}(\hat{\bf b}),  \bar{\sigma}^2 \mathbf{I}_{D\times D})$), where $D$ is the dimension of $\bf u$.
    Based on the above strategy, our objective function is expressed as
    \begin{align}\label{eq:obj}
        &\mathbb{E}_{{\bf{u}} \sim p_{\rm in}({\bf u})}\big[\mathbb{E}_{\hat{\bf{b}} \sim p_{\rm noisy}(\hat{\bf{b}}|\bf{u}; {\boldsymbol{\tilde{\mu}}}, {\boldsymbol{\tilde{d}}}, \theta)}[\log p_{\phi}(\bf{u}|\hat{\bf b})]\big]  \nonumber \\
        &\overset{(a)}{=} \mathbb{E}_{p_{\rm in}(\mathbf{u})}\left[\mathbb{E}_{p_{\rm noisy}(\hat{\bf{b}}|\bf{u}; {\boldsymbol{\tilde{\mu}}}, {\boldsymbol{\tilde{d}}}, \theta)} \left[\log\frac{\exp\big(-\frac{\|{\bf u}-f_{\phi}(\hat{\bf b})\|_2^2}{2\bar{\sigma}^2}\big)}{(2\pi\sigma^2)^{{D}/{2}}}\right]\right] \nonumber\\
       &= k - q\Big(\mathbb{E}_{p_{\rm in}(\mathbf{u})}\big[\mathbb{E}_{p_{\rm noisy}(\hat{\bf{b}}|\bf{u}; {\boldsymbol{\tilde{\mu}}}, {\boldsymbol{\tilde{d}}}, \theta)} [\|{\bf u}-f_{\phi}(\hat{\bf b})\|_2^2]\big]\Big),
    \end{align}
    where $(a)$ follows from the Gaussian assumption on the output of the decoder. 
    Since both $k$ and $q$ are positive constants, minimizing the mean squared error (MSE) between the input data $\bf u$ and the reconstructed data $\hat{\bf u}$ maximizes the original objective function in \eqref{eq:obj}. Therefore, we use the MSE loss function for maximizing the mutual information between $\bf{u}$ and $\hat{\bf{b}}$, defined as
    \begin{align}
        &\mathcal{L}_{\rm MSE} (\theta, \phi) 
        = \mathbb{E}_{{\bf u} \sim p_{\rm in}({\bf u})} \big[\mathbb{E}_{\hat{\bf b} \sim p_{\rm noisy}(\hat{\bf b}|{\bf u}; {\boldsymbol{\tilde{\mu}}}, \boldsymbol{\tilde{d}}, \theta)} [\|{\bf u}-f_{\phi}(\hat{\bf b}) \|_2^2]\big].     
        \label{MSE}
    \end{align}

    The ultimate objective of the digital semantic communication system in Sec.~II is to perform a dedicated machine learning task at the receiver. Motivated by this fact, we also introduce the loss function for maximizing the performance of the dedicated task by considering an image classification task as an example of such a task.
    For the loss function design, we aim at maximizing the mutual information between the true label $\ell({\bf u})$ corresponding to $\bf u$ and the reconstructed data $\hat{\bf u}$, in order to train the classifier $f_{\psi}$ to make accurate inference about the label using $\hat{\bf u}$. This can be regarded as essential information for task performance. 
    According to \cite{MI-LB}, to maximize $\mathit{I}(\ell({\bf u}); \bf{\hat u})$, it suffices to minimize the cross-entropy (CE) loss function defined as 
    \begin{align}
        &\mathcal{L}_{\rm CE} (\theta, \phi, \psi) \nonumber\\
        &=  -\mathbb{E}_{{\bf u} \sim p_{\rm in}({\bf u})} \big[ \mathbb{E}_{\hat{\bf b} \sim p_{\rm noisy}(\hat{\bf b}|{\bf u}; {\boldsymbol{\tilde{\mu}}}, \boldsymbol{\tilde{d}}, \theta)} \big[ \log f_{\psi}(f_{\phi}(\hat{\bf b}))_{\ell({\bf u})}\big] \big].
    \label{CE}
    \end{align}
    To maximize both reconstruction and classification accuracies, we finally design our loss function $\mathcal{L} (\theta, \phi, \psi)$ as the weighted summation of the MSE loss in \eqref{MSE} and the CE loss in \eqref{CE} as follows:
    \begin{align}
        \mathcal{L} (\theta, \phi, \psi)= \lambda \mathcal{L}_{\rm MSE} (\theta, \phi) + \mathcal{L}_{\rm CE} (\theta, \phi, \psi),
    \label{Class}
    \end{align}
    where $\lambda$ is a hyperparameter determined by the relative importance of the reconstruction accuracy of the JSCC encoder/decoder compared to the classification accuracy. In practice, when computing the loss, we replace the input distribution $p_{\rm in}({\bf u})$ using the empirical distribution obtained from a training dataset. 
    When executing gradient back-propagation, we exclude the quantization process after the JSCC encoder since this process involves non-differentiable sampling. Consequently, the computed gradients reach directly from the JSCC decoder's input $\hat{\bf b}$ to the JSCC encoder's output $f_{\theta}(\bf u)$.
    

    \subsection{Performance Enhancement via Warm-up Period}
    In the initial stage of the training process, the JSCC encoder faces challenges in generating meaningful latent variables due to its randomly initialized weights, which lack informative patterns. If these latent variables are further corrupted under the BSEC models with non-zero bit-flip and bit-erasure probabilities, the encoder may struggle to capture crucial input data features to maximize task performance. To address this challenge, we set the first $I_{\rm warmup}$ epochs of the training process as a warm-up period by setting $\tilde{\mu}_i=\tilde{d}_i =0, \forall i$, ensuring error-free transmission of latent variables. 
    Then, during this period, the weights of the JSCC encoder and decoder can be properly updated to generate informative latent variables to maximize the task performance without being disrupted by transmission errors. 
    Once the warm-up period ends, we assign non-zero bit-flip and bit-erasure probabilities to the BSEC models according to our sampling strategy in Sec.~\ref{Sec:Sample}. 
    In this subsequent period, the weights of the JSCC encoder and decoder are updated to enhance robustness against transmission errors while continuing to optimize the task performance.
 

    \section{Channel-Adaptive Modulation Technique of the Proposed JSCC Approach}\label{Sec:Modulation}
    In this section, we devise a channel-adaptive modulation technique for an inference phase, which reduces the communication latency of transmission while maintaining task performance at the receiver. 
    
    \subsection{BER Analysis for QAM}
    Our key observation is that training and testing environments are essentially the same if the bit-flip and bit-erasure probabilities of the QAM symbols transmitted during the inference phase are exactly the same as the parameters $\{\tilde{\mu}_i\}_{i=1}^N$ and $\{\tilde{d}_i\}_{i=1}^N$ of the BSECs set in the training phase. Motivated by this observation, we characterize the bit-flip and bit-erasure probabilities of the QAM symbols as a function of the channel condition and modulation order. 
    Let ${\sf BER}(M, {\sf SNR})$ be the BER of the $2^M$-QAM symbol for a given SNR.
    According to our demodulation method in \eqref{eq:robust_demod_final}, the bit error occurs when the real or imaginary part of the received signal passes the decision boundary that is close to the adjacent symbol, as illustrated in Fig.~\ref{fig:16QAM_detect}.
    This fact implies that the probability of a typical error event can be expressed as 
    \begin{align}
        \mathbb{P}\left(\mathfrak{Re}\left\{\tilde{v}[n]\right\} \geq \frac{(1+a_i)d_M}{2}\right)={\sf Q}\left((1+a_i)\sqrt{\frac{3{\sf SNR}}{2^M-1}}\right).
    \end{align}
    Based on the above result and the assumption of equiprobable bit
outputs from the JSCC encoder, the BER of the $2^M$-QAM symbol is approximately computed as 
    \begin{align}\label{eq:BER_final}
        {\sf BER}(M,{\sf SNR}) \approx \frac{4}{M}\left(1-\frac{1}{\sqrt{2^M}}\right){\sf Q}\left({(1+a_i)\sqrt{\frac{3{\sf SNR}}{2^M-1}}}\right),
    \end{align}
    where $4(1-1/\sqrt{2^M})/M$ is a factor to reflect the error event at the inner, edge, and corner in the constellation set \cite{SER_QAM}. 
    In a similar manner, the correct-decision probability of the $2^M$-QAM symbol is also computed as 
    \begin{align}
        1 -  \frac{4}{M}\left(1-\frac{1}{\sqrt{2^M}}\right){\sf Q}\left({(1-a_i)\sqrt{\frac{3{\sf SNR}}{2^M-1}}}\right).
    \end{align}
    By interpreting the bit-error and correct-decision probabilities as the parameters $\tilde{\mu}_i$ and $\tilde{r}_i$, respectively,  the bit-flip and bit-correct probabilities of the BSEC are characterized as
    \begin{align}\label{eq:flip_correct_final}
       \tilde{\mu}_i &= \frac{4}{M}\left(1-\frac{1}{\sqrt{2^M}}\right){\sf Q}\left({(1+a_i)\sqrt{\frac{3{\sf SNR}}{2^M-1}}}\right), \nonumber \\
        \tilde{r}_i &= 1 - \frac{4}{M}\left(1-\frac{1}{\sqrt{2^M}}\right){\sf Q}\left((1-a_i)\sqrt{\frac{3{\sf SNR}}{2^M-1}}\right), 
    \end{align}
    respectively.
    By utilizing the fact that $\tilde{d}_i = 1- \tilde{\mu}_i - \tilde{r}_i$, the bit-erasure probability of the BSEC is also determined by 
    \begin{align}
        \tilde{d}_i 
        &= \frac{4}{M}\left(1-\frac{1}{\sqrt{2^M}}\right) \left\{{\sf Q}\left((1-a_i)\sqrt{\frac{3{\sf SNR}}{2^M-1}}\right)  \right. \nonumber \\
        &\qquad\qquad\qquad\qquad~~~ \left. -~ {\sf Q}\left((1+a_i)\sqrt{\frac{3{\sf SNR}}{2^M-1}}\right) \right\}.\label{eq:erasure_final}
    \end{align}    

    \begin{figure}[t]
        \centering 
            {\epsfig{file=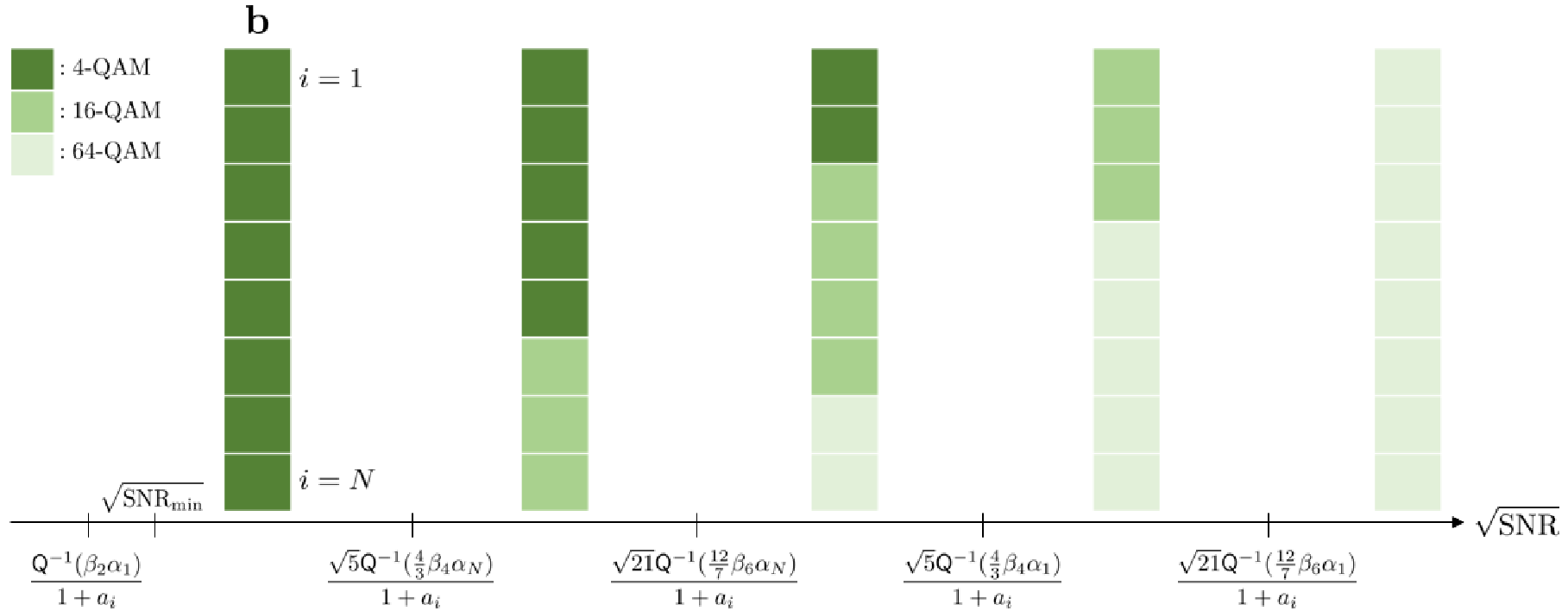, width=9cm}} 
        \caption{Changes in the modulation orders assigned to the latent variables for various SNRs when employing our channel-adaptive modulation technique.}
        \label{fig:ada_mod}
    \end{figure}
    
    \subsection{Channel-Adaptive Modulation}
    During the training phase, the bit-flip probability $\tilde{\mu}_i$ has been drawn from ${\rm Uniform}[0, \alpha_i]$, implying that the JSCC encoder and decoder are trained to cover the bit-flip probability up to the robustness level $\alpha_i$ which varies with $i$. Motivated by this fact, our channel-adaptive modulation technique chooses the highest modulation order that maintains the bit-error probability below a certain limit $\beta_{M}  \alpha_i$ (i.e., $\tilde{\mu}_i \leq \beta_{M} \alpha_i$), where $\beta_{M}  < 1$ is an adjusting factor to compensate for the effect of the BER approximation in \eqref{eq:BER_final}. 
    From the bit-error probability expression in \eqref{eq:BER_final}, our criterion for the $2^M$-QAM symbol is given by 
    \begin{align}
       \frac{4}{M}\left(1-\frac{1}{\sqrt{2^M}}\right){\sf Q}\left({(1+a_i)\sqrt{\frac{3{\sf SNR}}{2^M-1}}}\right)\leq \beta_{M} \alpha_i,
    \end{align}
    which is equivalent to the following SNR condition:
    \begin{align}\label{eq:SNR_condition}
       \sqrt{{\sf SNR}} \geq \underbrace{\frac{1}{1+a_i}\sqrt{\frac{2^M-1}{3}} {\sf Q}^{-1}\left(\frac{M\sqrt{2^M}}{4(\sqrt{2^M}-1)}\beta_{M} \alpha_i\right)}_{\triangleq \tau_{M,i}}.
    \end{align}
    The SNR condition in \eqref{eq:SNR_condition} implies that if the SNR falls below a certain threshold $\tau_{M,i}$, the modulation order $M$ must be reduced so that  the bit-error probability is sufficiently lower than a desired limit $\beta_{M} \alpha_i$.
    This fact aligns with our intuition because as the SNR decreases, lowering the modulation order becomes necessary to maintain a sufficiently low bit-error probability.
    Utilizing the SNR condition in \eqref{eq:SNR_condition},  the best modulation order to satisfy our criterion among three candidate orders\footnote{In this work, we only consider three modulation types, 4-QAM, 16-QAM, and 64-QAM, for simplicity; however, it is straightforward to extend the proposed technique to support a higher order modulation type, such as 256-QAM and 1024-QAM.}, $M\in\{2,4,6\}$, is determined as
    \begin{align}\label{diff_import_control_BSEC}
        M_i^\star &= \begin{cases}
        2, & \text{if}~~\tau_{2,i} \leq \sqrt{{\sf SNR}} \leq \tau_{4,i}, \\
        4, & \text{if}~~\tau_{4,i} \leq \sqrt{{\sf SNR}} \leq \tau_{6,i}, \\
        6, & \text{if}~~\tau_{6,i} \leq \sqrt{{\sf SNR}},  
        \end{cases} 
    \end{align}
    where
    $\tau_{2,i}=\frac{1}{1+a_i}{\sf Q}^{-1}(\beta_2\alpha_i)$, $\tau_{4,i}=\frac{1}{1+a_i}\sqrt{5}{\sf Q}^{-1}(\frac{4\beta_4\alpha_i}{3})$, and $\tau_{6,i}=\frac{1}{1+a_i}\sqrt{21}{\sf Q}^{-1}(\frac{12\beta_6\alpha_i}{7})$ from \eqref{eq:SNR_condition}. If the transmitter and receiver share the information $\{\alpha_i,a_i\}_i$ and $\{\beta_M\}_M$ as background knowledge, the receiver can compute the thresholds $\{\tau_{M,i}\}_{M,i}$. Then, based on \eqref{diff_import_control_BSEC}, the receiver can determine the modulation orders of the binary latent variables without requiring explicit information exchanges.
    It should be noted that we properly set the robustness level $\alpha_i$ so that every latent variable is modulated as the $4$-QAM symbol (i.e., $M_i^\star \geq 2$) even in the worst SNR case.
    To provide more insights about our technique, in Fig.~\ref{fig:ada_mod}, we illustrate how the modulation orders assigned to the latent variables change as the SNR increases, when employing our technique. In this figure, $\beta_2$, $\beta_4$, and $\beta_6$ are set as assumed in the heterogeneous setting described in Sec.~\ref{Sec:Simul}.
    As can be seen in Fig.~\ref{fig:ada_mod}, the modulation types constituting the symbol sequence change in the following order: 
    \begin{align}
        &{\rm 4 QAM} \rightarrow ({\rm 4 QAM}, {\rm 16 QAM}) \rightarrow ({\rm 4 QAM}, {\rm 16 QAM}, {\rm 64 QAM}) \nonumber\\
        &\rightarrow ({\rm 16 QAM}, {\rm 64 QAM}) \rightarrow {\rm 64 QAM}, \nonumber 
    \end{align}
    as the SNR increases.
    Fig.~\ref{fig:ada_mod} clearly demonstrates that the better the channel condition, the higher the modulation order chosen by each latent variable.
    It is also shown that different modulation orders are assigned across the latent variables according to their robustness levels. 

    A key feature of our adaptive modulation technique is that it allows digital semantic communication to adapt not only to the instantaneous channel condition during the inference phase but also to the robustness levels of the latent variables chosen during the training phase. Another key feature is that when employing our technique, the modulation orders assigned to the latent variables increase as the SNR increases. This result coincides with our intuition because when the SNR is sufficiently high, the use of high-order modulation improves the overall communication latency of the system while maintaining the bit-error probability to be less than an acceptable level. Therefore, the use of our adaptive modulation technique provides flexibility in the average spectral efficiency over a wide range of SNR values, enabling our technique to adaptively minimize the communication latency according to the SNR while maintaining task performance.

    \vspace{3mm}
    {\bf Remark 2 (Comparison with Conventional AMC Technique):}
    Our channel-adaptive modulation technique resembles with a conventional adaptive modulation and coding (AMC) technique that has been widely adopted in modern wireless standards. 
    In this technique, both modulation order and coding rate are adaptively determined according to the SNR, in order to maximize the spectral efficiency while ensuring a sufficiently low block error rate. 
    A key feature that distinguishes our technique from the conventional AMC technique is its ability to assign different modulation orders across data bits (i.e., binary latent variables) even under the same SNR.
    This flexibility arises from assigning diverse robustness levels to the latent variables during the training phase, enhancing the adaptability of modulation orders, as detailed in Sec.~\ref{Sec:Sample}.
    Therefore, our technique can be viewed as a judicious extension of the conventional AMC technique, strategically designed to minimize the communication latency of the digital semantic communication system while maintaining task performance at the receiver. The performance gain achieved by assigning different modulation orders will be numerically demonstrated in Sec.~\ref{Sec:Simul}.

   \section{Simulation Results}\label{Sec:Simul}
   In this section, we evaluate the superiority of the proposed JSCC approach through simulations using the MNIST \cite{MNIST}, Fashion-MNIST \cite{Fashion-MNIST}, CIFAR-10 and CIFAR-100 \cite{CIFAR-10} datasets. The MNIST and Fashion-MNIST datasets consist of 60,000 training images and 10,000 test images. The CIFAR-10 and CIFAR-100 datasets consist of 50,000 training images and 10,000 test images. We normalize the training and test data to have a zero mean and unit variance \cite{pytorch}. The input image size for the JSCC encoder is $(1, 28, 28)$ for MNIST and Fashion-MNIST and $(3, 32, 32)$ for CIFAR-10 and CIFAR-100. 
   When training the JSCC encoder and decoder, we use the loss defined in \eqref{Class} for an image classification task, the MSE loss in \eqref{MSE} for an image reconstruction task, and the CE loss in \eqref{CE} for an image retrieval task. For the loss in \eqref{Class}, we set $\lambda = 0.2$. For CIFAR-10, $N$ is set to 128 for the image classification and retrieval task and 512 for the image reconstruction task. An Adam optimizer \cite{Adam} with a learning rate of 0.001 is employed for all the datasets. The batch size is set as $|\mathcal{D}| =256$. The total number of epochs is $30$ for MNIST and Fashion-MNIST, $20$ for CIFAR-10, and $500$ for CIFAR-100. Table~\ref{table:1} summarizes the neural network architectures for the considered datasets, where DO$(p)$ indicates that a dropout strategy is applied with a probability of $p$, BN stands for batch normalization, LR represents LeakyReLU with a slope of 0.2, and MP denotes max-pooling with kernel size $(2\times2)$.

   In our simulations, we consider the following approaches for performance evaluation:
   \begin{table}[t]
        \renewcommand{\arraystretch}{1.2}
        \caption{The model structure for image reconstruction, image classification and image retrieval tasks on MNIST, Fashion-MNIST, CIFAR-10 and CIFAR-100 datasets.}\label{table:model}
        \setlength{\tabcolsep}{3pt}
        \footnotesize 
        \centering
        \begin{adjustbox}{max width=0.45\textwidth}
        {\begin{tabular}{|c|c|c|c|} \hline
            \multicolumn{2}{|c|}{} & \multicolumn{1}{c|}{Layers} & \multicolumn{1}{c|}{Output size}   \\ \hline \hline 

            & \multirow{3}{*}{Encoder} & \multicolumn{1}{c|}{Dense+ReLU} & \multicolumn{1}{c|}{512} \\
            & & \multicolumn{1}{c|}{Dense+ReLU} & \multicolumn{1}{c|}{256} \\
            & & \multicolumn{1}{c|}{Dense+Sigmoid} & \multicolumn{1}{c|}{$N=96$} \\ \cline{2-4} 
            
            & \multirow{3}{*}{Decoder} & \multicolumn{1}{c|}{Dense+ReLU} & \multicolumn{1}{c|}{256} \\
            \multirow{6}{*}{Fashion-MNIST} & & \multicolumn{1}{c|}{Dense+ReLU} & \multicolumn{1}{c|}{512} \\
            MNIST/ & & \multicolumn{1}{c|}{Dense+Tanh} & \multicolumn{1}{c|}{784} \\ \cline{2-4} 
            
            & \multirow{6}{*}{Classifier} & \multicolumn{1}{c|}{Dense+ReLU} & \multicolumn{1}{c|}{256} \\
            & & \multicolumn{1}{c|}{Dense+ReLU} & \multicolumn{1}{c|}{256} \\
            & & \multicolumn{1}{c|}{Dense+ReLU} & \multicolumn{1}{c|}{256} \\ 
            & & \multicolumn{1}{c|}{Dense+ReLU} & \multicolumn{1}{c|}{128} \\ 
            & & \multicolumn{1}{c|}{Dense+ReLU+DO(0.5)} & \multicolumn{1}{c|}{128} \\ 
            & & \multicolumn{1}{c|}{Dense} & \multicolumn{1}{c|}{10} \\ \hline \hline

            \multirow{14}{*}{CIFAR-10}  & \multirow{3}{*}{Encoder} & \multicolumn{1}{c|}{Table 1 in \cite{revision_ref_5}} & \multicolumn{1}{c|}{$(8,8,8)$} \\
            & & \multicolumn{1}{c|}{Flatten} & \multicolumn{1}{c|}{$512$} \\
            & & \multicolumn{1}{c|}{Dense+Sigmoid} & \multicolumn{1}{c|}{$N$} \\ \cline{2-4} 

            & \multirow{3}{*}{Decoder} & \multicolumn{1}{c|}{Dense} & \multicolumn{1}{c|}{$512$} \\
            & & \multicolumn{1}{c|}{Unflatten} & \multicolumn{1}{c|}{$(8,8,8)$} \\
            & & \multicolumn{1}{c|}{Table 1 in \cite{revision_ref_5}} & \multicolumn{1}{c|}{$(3,32,32)$} \\ \cline{2-4} 
            
            & \multirow{8}{*}{Classifier} & \multicolumn{1}{c|}{3x3Conv+BN+LR} & \multicolumn{1}{c|}{$(32,32,32)$} \\
            & & \multicolumn{1}{c|}{3x3Conv+BN+LR+MP+DO(0.25)} & \multicolumn{1}{c|}{$(32,16,16)$} \\
            & & \multicolumn{1}{c|}{3x3Conv+BN+LR} & \multicolumn{1}{c|}{$(64,16,16)$} \\ 
            & & \multicolumn{1}{c|}{3x3Conv+BN+LR+MP+DO(0.25)} & \multicolumn{1}{c|}{$(64,8,8)$} \\ 
            & & \multicolumn{1}{c|}{3x3Conv+BN+LR} & \multicolumn{1}{c|}{$(128,8,8)$} \\ 
            & & \multicolumn{1}{c|}{3x3Conv+BN+LR+MP+DO(0.25)} & \multicolumn{1}{c|}{$(128,4,4)$} \\
            & & \multicolumn{1}{c|}{Dense+LR+DO(0.25)} & \multicolumn{1}{c|}{128} \\
            & & \multicolumn{1}{c|}{Dense} & \multicolumn{1}{c|}{10} \\ \hline \hline 

            \multirow{7}{*}{CIFAR-100}  & \multirow{3}{*}{Encoder} & \multicolumn{1}{c|}{Table 2 in \cite{revision_ref_5} + Remove one residual block} & \multicolumn{1}{c|}{$(8,8,8)$} \\
            & & \multicolumn{1}{c|}{Flatten} & \multicolumn{1}{c|}{$512$} \\
            & & \multicolumn{1}{c|}{Dense+Sigmoid} & \multicolumn{1}{c|}{$N=128$} \\ \cline{2-4} 

            & \multirow{4}{*}{Retrieval} & \multicolumn{1}{c|}{Dense} & \multicolumn{1}{c|}{$512$} \\
            & & \multicolumn{1}{c|}{Unflatten} & \multicolumn{1}{c|}{$(8,8,8)$} \\
            & & \multicolumn{1}{c|}{Table 2 in \cite{revision_ref_5} + Remove one residual block} & \multicolumn{1}{c|}{$(3,32,32)$} \\ 
             & & \multicolumn{1}{c|}{Resnet-18 in \cite{revision_ref_6}} & \multicolumn{1}{c|}{$20$} \\ \hline 

        \end{tabular}} 
        \end{adjustbox}
        \label{table:1}
    \end{table} 

    \begin{figure*}
        \centering
        \begin{minipage}{2\columnwidth}
            \centering
            \subfigure[MNIST]
            {\epsfig{file=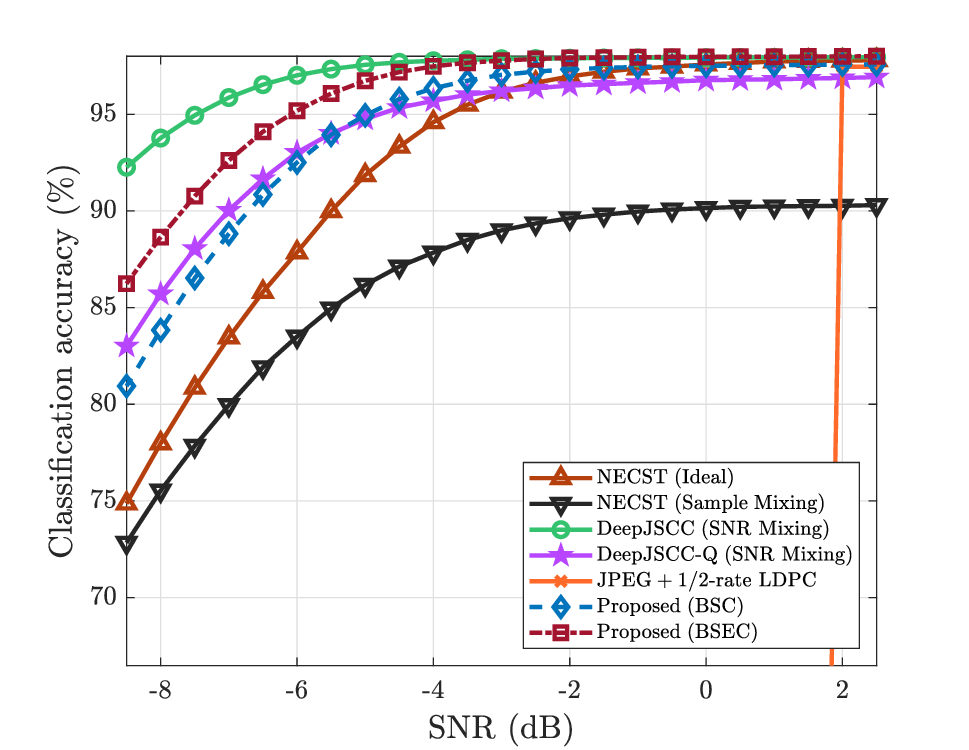, width=6cm}}
            \subfigure[Fashion-MNIST]
		  {\epsfig{file=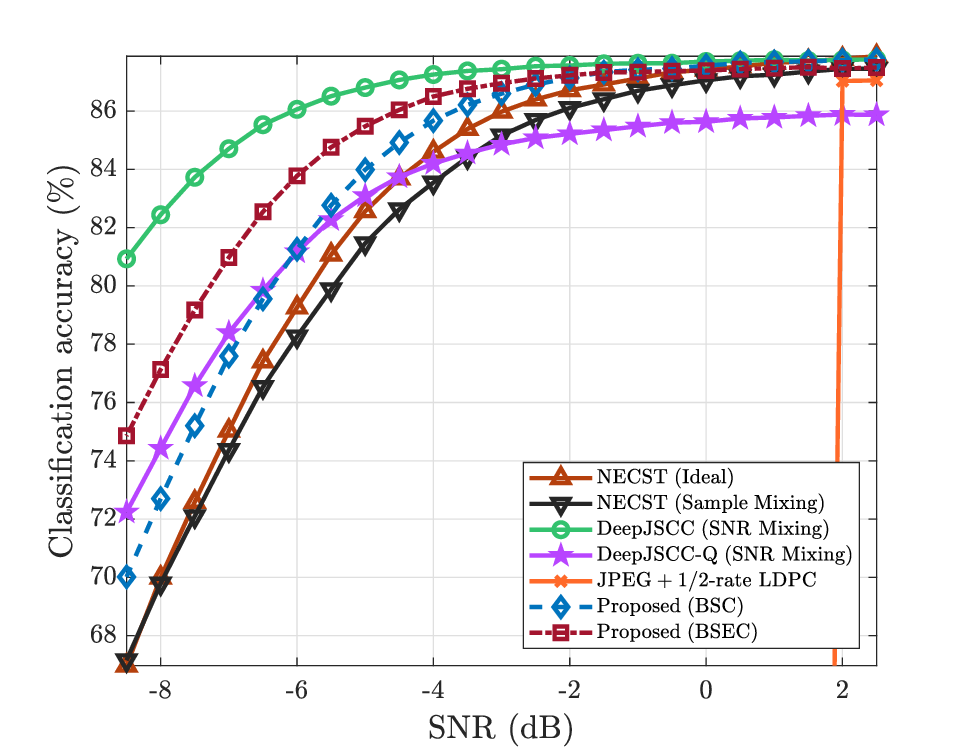, width=6cm}}
            \subfigure[CIFAR-10]
		  {\epsfig{file=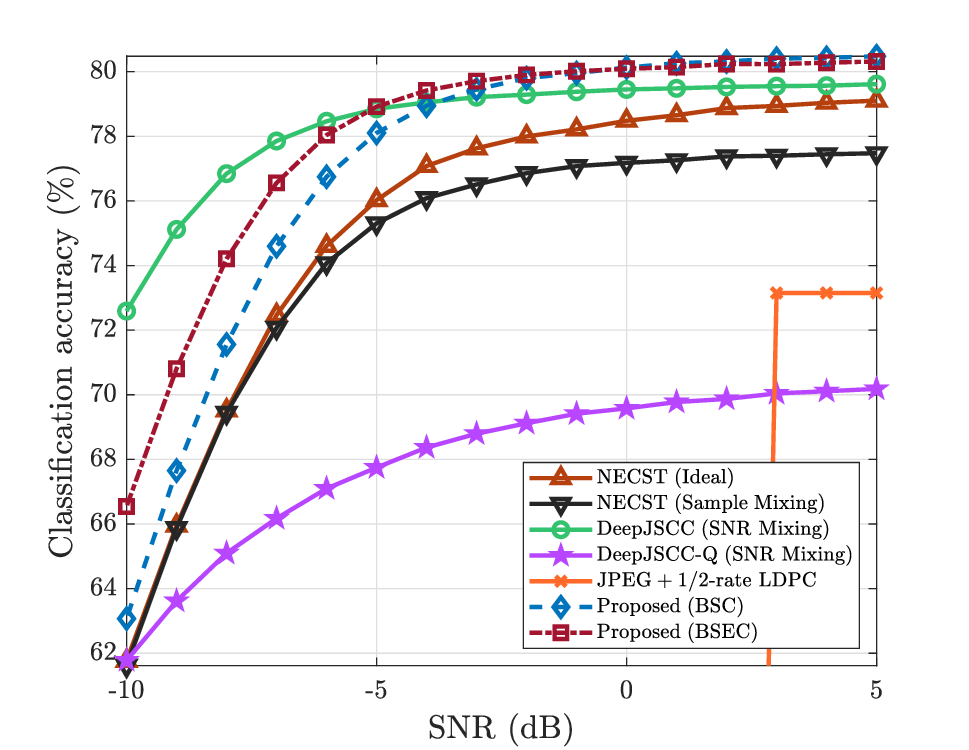, width=6cm}}
            \captionof{figure}{Comparison of the classification accuracies of various JSCC and non-JSCC approaches with 4-QAM for image classification tasks using the MNIST, Fashion-MNIST, and CIFAR-10 datasets.}
            \label{fig:class_acc}
        \end{minipage}
    \end{figure*}

   \begin{itemize}
    \item {\bf Proposed}: We consider two variants of the proposed JSCC approach with different demodulation methods: (i) {\bf Proposed (BSEC)}, utilizing a robust demodulation method with $a_i=0.5$, $\forall i$, modeled by the BSECs during training, and (ii) {\bf Proposed (BSC)}, employing conventional hard-output demodulation with $a_i=0$, $\forall i$, modeled by the BSCs during training.
    The number of the epochs for the warm-up period is set as $I_{\rm warmup}=5$.
    When employing fixed modulation, we set $\alpha_i=0.45$, $\forall i$, in {\bf Proposed (BSC)} and $\alpha_i=0.4$, $\forall i$, in {\bf Proposed (BSEC)}, unless specified otherwise.
    When employing our channel-adaptive modulation technique, we consider two scenarios, referred to as  {\em homogeneous} and {\em heterogeneous}.
    In the homogeneous setting, we set $\alpha_i=0.4$, $\forall i$, and set $(\beta_2,\beta_4,\beta_6) = (0.6599,0.6003,0.5553)$. Note that in this setting, $\beta_M$ is properly selected to ensure that the higher modulation order is chosen for higher SNR values (i.e., $\tau_{2,i} \leq \tau_{4,i} \leq \tau_{6,i}$, $\forall i$). 
    In the heterogeneous setting, we set $\alpha_i = \frac{\alpha_N-\alpha_1}{N-1}(i-1)+\alpha_1$ with $\alpha_1 = 0.29$ and $\alpha_N= 0.45$, and also set $(\beta_2,\beta_4,\beta_6) = (1,0.6,0.5)$ to satisfy both  $\tau_{2,i} \leq \tau_{4,i} \leq \tau_{6,i}$, $\forall i$, and $\tau_{2,1} \leq \tau_{4,N} \leq \tau_{6,N} \leq \tau_{4,1} \leq \tau_{6,1}$.
    


    \item {\bf NECST (Ideal)}: We consider the neural joint source-channel coding (NECST) approach in \cite{NECST}, which assumes the BSCs with the homogeneous bit-flip probabilities, with an {\em ideal} training strategy described below. In this strategy, we consider multiple pairs of JSCC encoders and decoders and then train each pair of the JSCC encoder and decoder for a specific SNR value range, which is of interest during the inference phase.
    During the inference phase, every time the system encounters the channel condition with a particular SNR, the system selects the pair of the JSCC encoder and decoder trained with the corresponding SNR. This strategy provides the best performance for the NECST approach in \cite{NECST}, but requires a large number of pairs of the JSCC encoder and decoder trained with various SNR values. 

    \item {\bf NECST (Sample Mixing)}: We also consider the NECST approach in \cite{NECST} with a {\em sample mixing} training strategy described below. In this strategy, we divide the batch into 8 sub-batches of equal size. When training with each sub-batch, we choose one of the various bit-flip probabilities for the BSC modeling. For example: 0.01, 0.01, 0.05, 0.1, 0.15, 0.2, 0.25, 0.3, 0.35. This approach enables a single pair of the JSCC encoder and decoder to be trained for a wide range of SNR values. 
    
    \item {\bf DeepJSCC}: We modify an analog JSCC approach considered in \cite{DeepJSCC} using our JSCC encoder and decoder structure, while training the model through a stochastic approach with $I_{\rm warmup} = 5$. In particular, to ensure the average symbol power is the same as normalized QAM symbols, we multiply a factor $2$ to the JSCC encoder, ensuring that its output has a value between $0$ and $2$, closely resembling an average power of 1. 
    During the training process in {\bf DeepJSCC}, the SNR is sampled from the uniform distribution, where the minimum and maximum values correspond to the ends of the SNR range of interest during the inference phase. For the image retrieval task, we set the number of epochs to 200, in order to avoid an overfitting problem.

    \item {\bf DeepJSCC-Q}: We modify a digital JSCC approach considered in \cite{DeepJSCC-Q} using our JSCC encoder and decoder structure, while training the model through a stochastic approach with $I_{\rm warmup} = 5$. This approach conducted end-to-end training by incorporating the nonlinearity of digital modulation. 
    In particular, to ensure the average symbol power is the same as normalized QAM symbols, we remove the sigmoid in the last layer of the encoder and performed power normalization. During the training process in {\bf DeepJSCC-Q}, the SNR is sampled from the uniform distribution, where the minimum and maximum values correspond to the ends of the SNR range of interest during the inference phase. Unlike other approaches, a learning rate is set as 0.00005 for {\bf DeepJSCC-Q} as it provides the best performance.

    \item {\bf JPEG+1/2-rate LDPC}: We consider a separate source-channel coding approach employing JPEG for source coding and 1/2-rate LDPC for channel coding. For the image classification task, a noise-free classifier is separately trained with a learning rate of 0.00005 with the batch size of $|\mathcal{D}|=32$.
    
   \end{itemize}

    \begin{figure}[t]
        \centering 
            {\epsfig{file=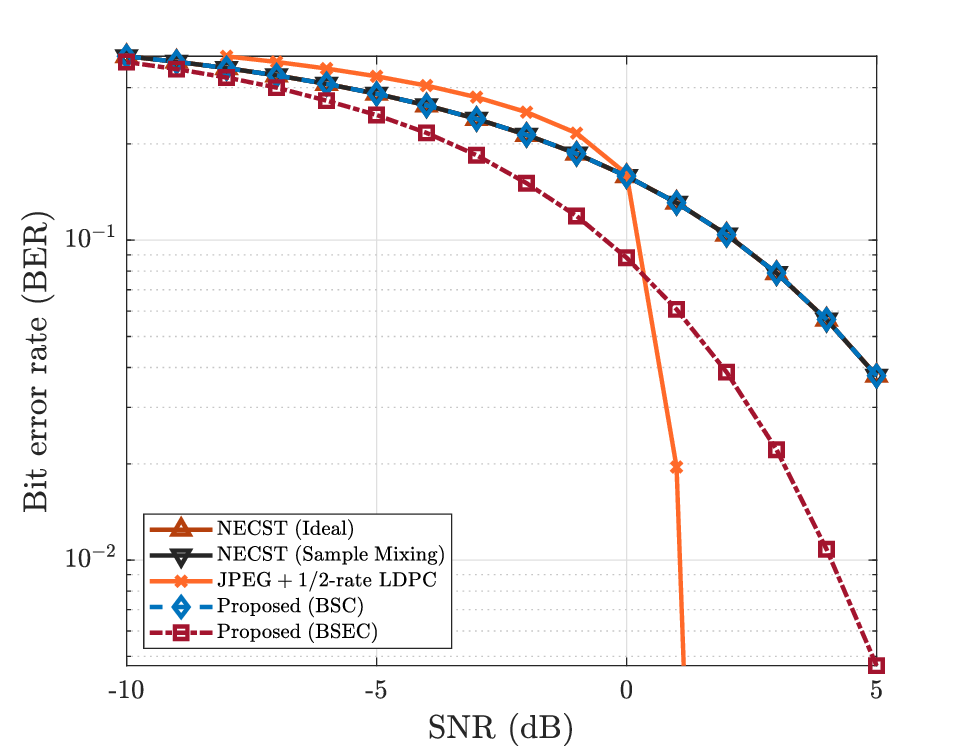, width=7.4cm}} 
        \caption{Comparison of the BERs of various JSCC and non-JSCC approaches with 4-QAM for the image classification task using the CIFAR-10 dataset.}
        \label{fig:ber_class}
    \end{figure}

    \subsection{Performance Evaluation with Fixed Modulation}

    Fig.~\ref{fig:class_acc} compares the classification accuracies of various JSCC and non-JSCC approaches with 4-QAM for image classification tasks using the MNIST, Fashion-MNIST, and CIFAR-10 datasets. 
    For {\bf JPEG+1/2-rate LDPC}, the average bit sequence length $N$ is 9282, 10646, and 14966 for the MNIST, Fashion-MNIST, and CIFAR-10 datasets, respectively.
    Fig.~\ref{fig:class_acc} shows that the proposed approaches achieve a higher classification accuracy than the existing NECST approaches, particularly in the low-SNR regime. 
    In particular, {\bf Proposed (BSEC)} using a single encoder-decoder pair even outperforms {\bf NECST (Ideal)} which requires a large number of the encoder-decoder pairs. This result validates the effectiveness of our robust training strategy in Sec.~\ref{Sec:Training} in improving the robustness of the JSCC encoder and decoder against diverse SNR. It is also shown that {\bf Proposed (BSEC)} which adopts the robust demodulation method in Sec.~\ref{Sec:Demod} outperforms {\bf Proposed (BSC)} which utilizes the conventional hard-output demodulation. This result demonstrates the superiority of our demodulation method over the conventional method. Although {\bf DeepJSCC} exhibits the highest classification accuracy in some low-SNR regimes, {\bf DeepJSCC} is not compatible with practical digital communication systems. In addition, {\bf DeepJSCC} has a longer communication latency compared to other JSCC approaches because {\bf DeepJSCC} requires $N$ channel uses to transmit $N$ real-values. 
    {\bf Proposed (BSEC)} also outperforms {\bf DeepJSCC-Q} in terms of the classification accuracy for all the datasets. This performance gain can be attributed to the utilization of the demodulation process, which maps real-domain equalized signals into ternary values. The statistical modeling and robust training based on demodulation enable the JSCC decoder to effectively manage errors in the transmission of latent variables. The performance of {\bf JPEG+1/2-rate LDPC} degrades significantly below a certain SNR threshold, despite requiring a bit sequence length more than 97 times longer than other JSCC approaches, including {\bf Proposed (BSEC)}. This result demonstrates the superiority of the proposed JSCC approach over a traditional separate source-channel coding approach in terms of both communication latency and task performance.  



    \begin{figure*}[t]
        \begin{minipage}{2\columnwidth}
            \centering
            \subfigure[PSNR, CIFAR-10]
            {\epsfig{file=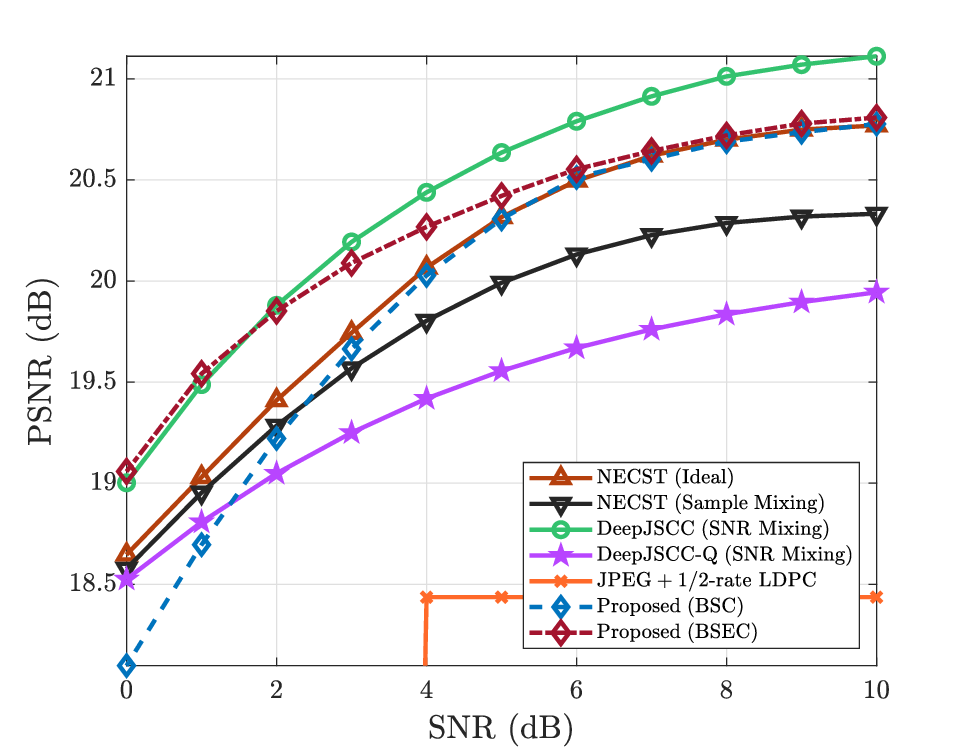, width=7.4cm}}
            \hspace{3mm}
            \subfigure[mAP, CIFAR-100]
		{\epsfig{file=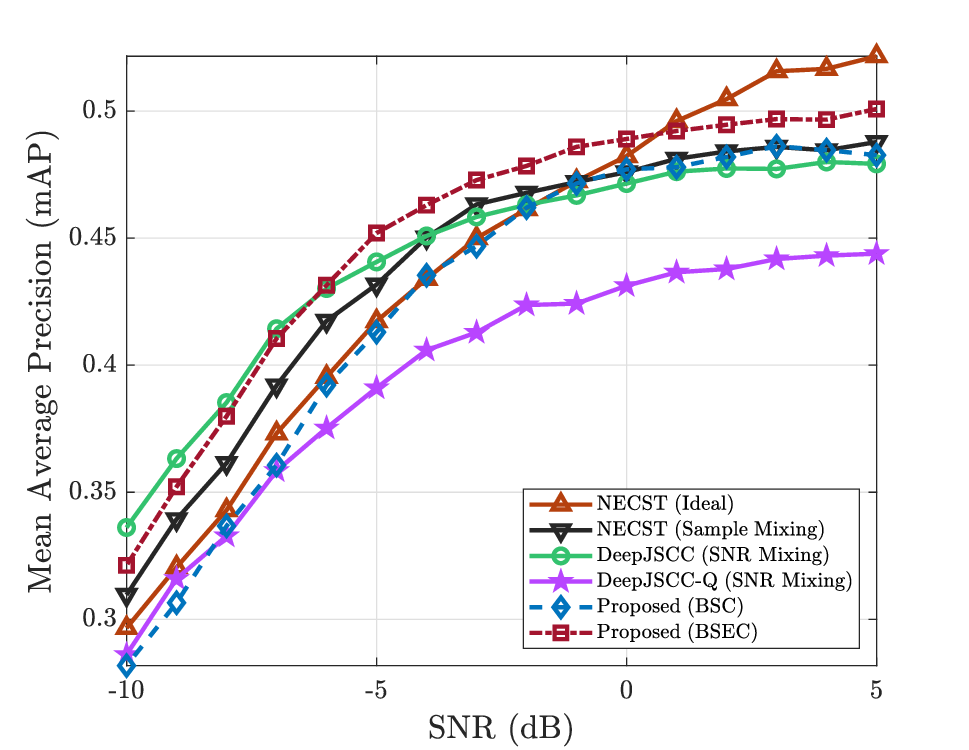, width=7.4cm}} 
            \captionof{figure}{
            Comparison of the PSNRs and mAPs of various approaches with 4-QAM for the image reconstruction and retrieval tasks using the CIFAR-10 and CIFAR-100 datasets.}
           \label{fig:PSNR_mAP}
        \end{minipage}
    \end{figure*}
    
    Fig.~\ref{fig:ber_class} compares the BERs of various JSCC and non-JSCC approaches for the image classification task using the CIFAR-10 dataset. The BER is computed by counting the number of bit-flip errors. The BERs of {\bf NECST (Ideal)}, {\bf NECST (Sample Mixing)}, and {\bf Proposed (BSC)} are exactly the same because these methods employ the same hard-output demodulation. {\bf Proposed (BSEC)} outperforms these methods in terms of the BER, which is achieved by employing the proposed robust demodulation. This result demonstrates that the proposed demodulation method effectively mitigates frequent bit-flip errors. Although {\bf JPEG+1/2-rate LDPC} achieves the lowest BER when ${\sf SNR} > 1$ dB, it suffers from significant degradation in performance when ${\sf SNR} < 1$ dB, resulting in decoding failure. Consequently, the recovery of the semantic information at the receiver completely fails, as evidenced by the results in Fig.~\ref{fig:class_acc}. These results demonstrate the unsuitability of the conventional non-JSCC approaches for semantic communications in low-SNR regimes. 

    In {\bf NECST} or {\bf Proposed (BSC)}, conventional hard-output demodulation is performed before JSCC decoding. The computational complexity required for this demodulation is of the order $\mathcal{O}(2 \cdot \frac{N}{M} \cdot 2^{M/2})$. In contrast, the {\bf Proposed (BSEC)} method employs a robust demodulation complexity denoted as $\mathcal{O}(3 \cdot \frac{N}{M} \cdot 2^{M/2})$. On the other hand, {\bf DeepJSCC} or {\bf DeepJSCC-Q} directly input the equalized received signal to the JSCC decoder, eliminating the need for a separate demodulation step. These comparisons reveal that {\bf Proposed (BSEC)} offers performance gains at the expense of additional complexity compared to existing JSCC approaches. However, this additional complexity remains comparable to the typical demodulation process in traditional communication systems and is therefore acceptable for practical implementation.


     Fig.~\ref{fig:PSNR_mAP}(a) compares the peak signal-to-noise ratios (PSNRs) of various JSCC and non-JSCC approaches with 4-QAM for the image reconstruction task using the CIFAR-10 dataset. In this simulation, data normalization is not executed, while the activation function of the final layer of the JSCC decoder is replaced with a sigmoid function. For both {\bf Proposed (BSEC)} and {\bf Proposed (BSC)}, we set $\alpha_i=0.05$, $\forall i$. For {\bf JPEG+1/2-rate LDPC}, the average bit sequence length $N$ is set as 10906. Fig.~\ref{fig:PSNR_mAP}(a) shows that {\bf Proposed (BSEC)} outperforms other JSCC approaches in terms of PSNR except {\bf DeepJSCC}. Although {\bf DeepJSCC} exhibits a better image reconstruction quality than the proposed approaches when ${\sf SNR} \geq \rm 3$ dB, {\bf DeepJSCC} has no compatibility with practical digital systems and requires a longer communication latency than other JSCC approaches. {\bf Proposed (BSEC)} exhibits a significant performance improvement over {\bf Proposed (BSC)}, affirming the efficacy of our robust demodulation method combined with the BSEC modeling approach. 
     {\bf JPEG+1/2-rate LDPC} is inferior to {\bf Proposed (BSEC)}, while the bit sequence length of {\bf JPEG+1/2-rate LDPC} is approximately 21 times larger than that of other JSCC approaches. Therefore, our results in Fig.~\ref{fig:PSNR_mAP}(a) demonstrate the advantage of the JSCC approaches over a traditional separate source-channel coding approach in terms of both communication latency and reconstruction performance, as already observed in Fig.~\ref{fig:class_acc}.
     
    
     Fig.~\ref{fig:PSNR_mAP}(b) compares the  mean average precisions (mAPs) of various JSCC approaches with 4-QAM for the image retrieval task  using the CIFAR-100 dataset. Note that the mAP is a well-known performance metric for measuring image retrieval accuracy \cite{mAP}. 
     During retrieval, image searching is conducted until three images belonging to the same class as the query image is found. The mAP is calculated for 20 superclasses in the CIFAR-100 dataset. Fig.~\ref{fig:PSNR_mAP}(b) shows that {\bf Proposed (BSEC)} outperforms other digital-based JSCC approaches at $\sf SNR \leq 0~dB$. Additionally, at $\sf SNR = 5~dB$, {\bf Proposed (BSEC)}, which requires only a single encoder-decoder pair, exhibits a marginal difference of 2.08$\%$ compared to {\bf NECST (Ideal)}, which requires a large number of encoder-decoder pairs. {\bf Proposed (BSEC)} exhibits performance comparable to {\bf DeepJSCC} at $\sf SNR \leq -3~dB$. These results verify the superiority of the proposed approach over the existing JSCC approaches even for the image retrieval task.

    \subsection{Performance Evaluation with Channel-Adaptive Modulation}
    \begin{figure}[t]
        \centering 
            {\epsfig{file=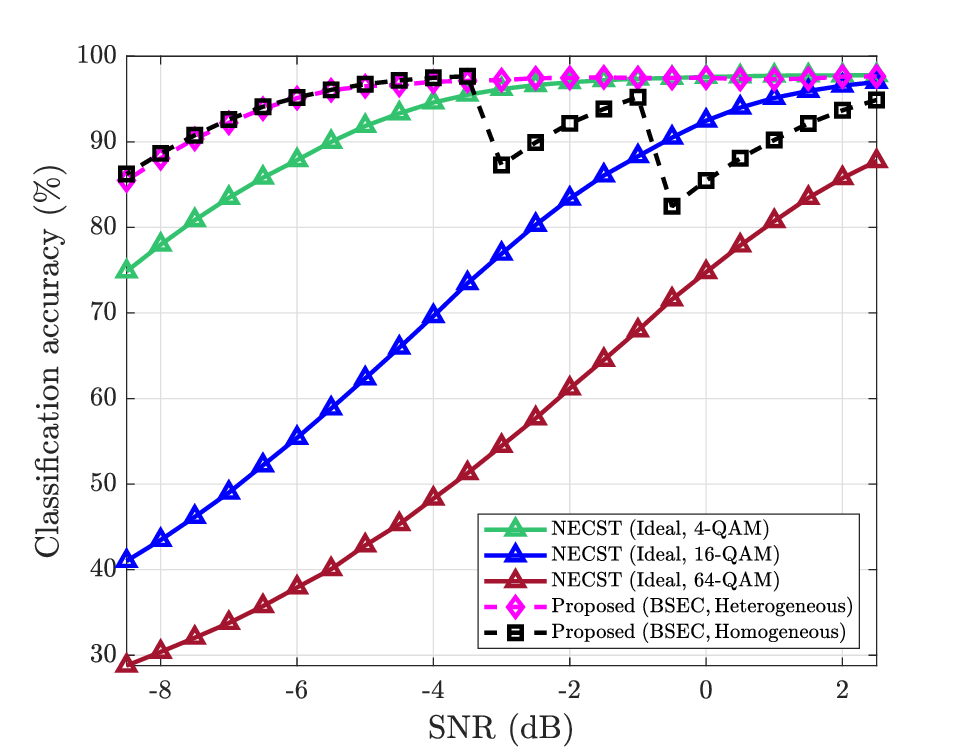, width=7.4cm}} 
        \caption{Comparison of the classification accuracies of the proposed JSCC approach with the channel-adaptive modulation and the NECST approach with a fixed modulation for an image classification task on the MNIST dataset.}
        \label{fig:ada_mod_control_1}
    \end{figure}

    \begin{figure*}[t]
        \begin{minipage}{2\columnwidth}
            \centering
            \subfigure[Classification accuracy]
            {\epsfig{file=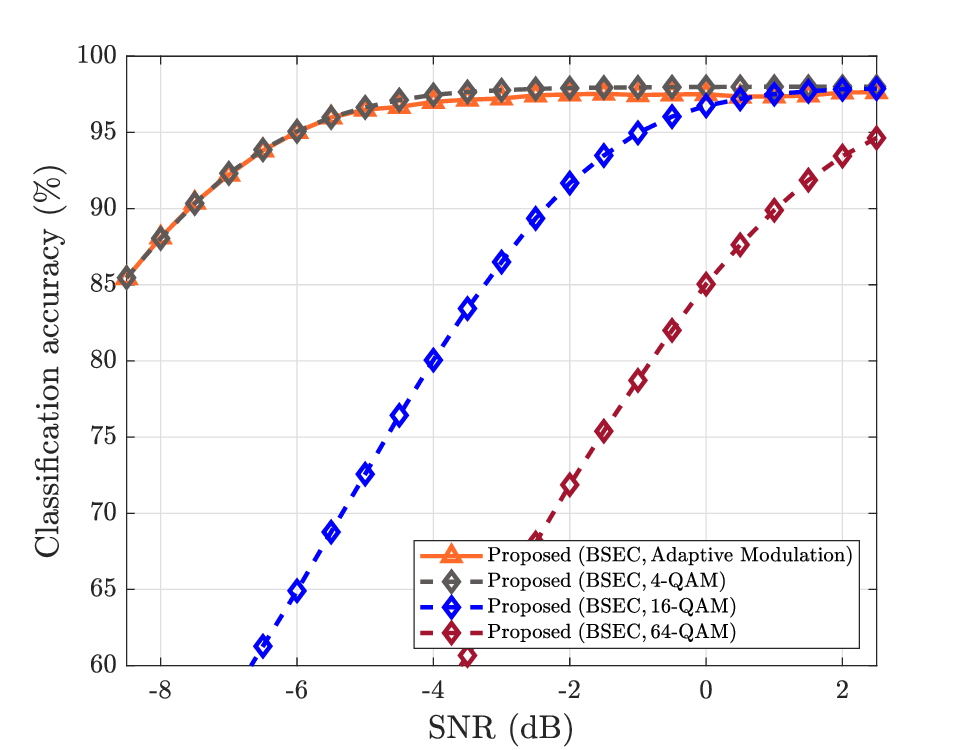, width=7.4cm}}
            \hspace{3mm}
            \subfigure[Spectral efficiency]
		{\epsfig{file=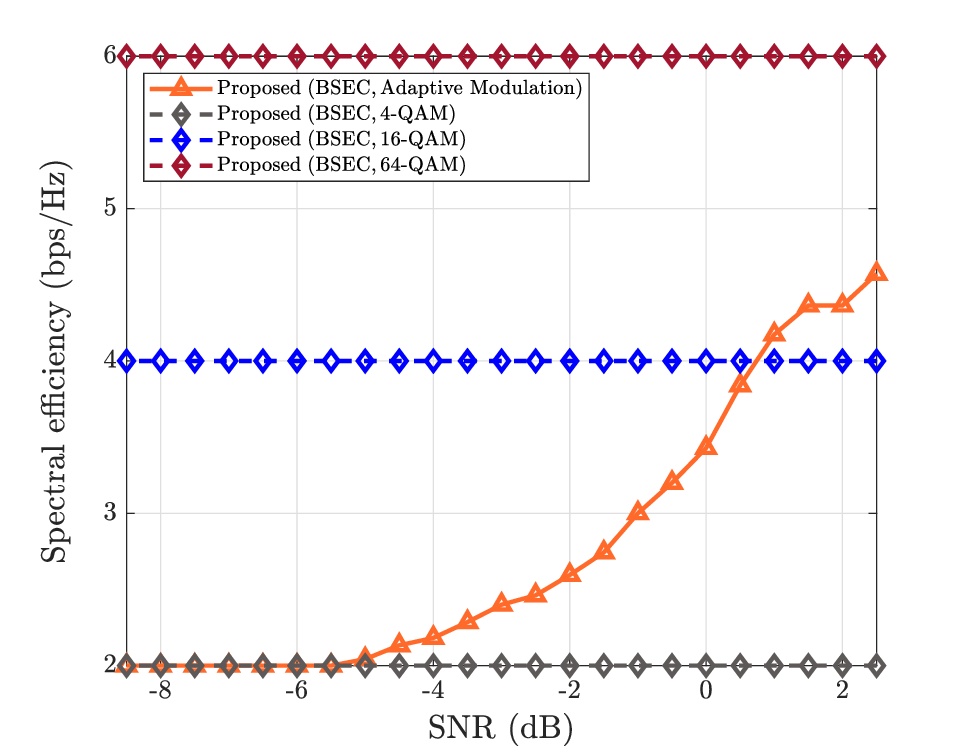, width=7.4cm}} 
            \captionof{figure}{
            Comparison of the classification accuracy and spectral efficiency of the proposed JSCC approach with and without the channel-adaptive modulation for an image classification task on the MNIST dataset.}
           \label{fig:SE}
        \end{minipage}
    \end{figure*}
    Fig.~\ref{fig:ada_mod_control_1} compares the classification accuracies of the proposed JSCC approach with the channel-adaptive modulation and the NECST approach with a fixed modulation type for an image classification task on the MNIST dataset.
    Fig.~\ref{fig:ada_mod_control_1} shows that {\bf Proposed (BSEC)} with the heterogeneous setting outperforms {\bf NECST (Ideal)} with a fixed modulation type for the entire SNR regime.
    These results demonstrate the effectiveness of the channel-adaptive modulation technique, enabling the adaptive selection of modulation orders based on varying SNR and robustness levels.
    Fig.~\ref{fig:ada_mod_control_1} also shows that {\bf Proposed (BSEC)} with the homogeneous setting suffers from performance loss at higher modulation orders. 
    This performance degradation arises from the use of the same modulation order across all latent variables, occurring when employing the same robustness level, as indicated in \eqref{diff_import_control_BSEC}.  
    For example, when $-3\text{ dB} \leq {\sf SNR} \leq -1\text{ dB}$, {\bf Proposed (BSEC)} with the homogeneous setting chooses 16-QAM, resulting in the inferior performance compared to the NECST approach with 4-QAM. A similar result is observed when $-0.5\text{ dB} \leq {\sf SNR} \leq 2.5\text{ dB}$, where {\bf Proposed (BSEC)} with the homogeneous setting chooses 64-QAM.
    Our results clearly highlight that assigning different robustness levels across the latent variables is essential for the proposed JSCC approach not only to ensure flexibility in selecting modulation orders, but also to maximize task performance.
    Nevertheless, the proposed approach with the homogeneous setting still outperforms {\bf NECST (Ideal)} when both approaches utilize the same modulation type.

    Fig.~\ref{fig:SE} illustrates the classification accuracy and spectral efficiency of the proposed JSCC approach with and without the channel-adaptive modulation for an image classification task on the MNIST dataset.
    In this simulation, we adopt the heterogeneous setting for the proposed approach.
    Fig.~\ref{fig:SE}(a) shows that the classification accuracy of {\bf Proposed (BSEC)} with a fixed modulation decreases with the modulation order, particularly in the low SNR regime. 
    This is because the bit-error probability decreases as the modulation order decreases, leading to the fundamental trade-off between the classification accuracy and spectral efficiency.   
    Although this trade-off is inevitable, the classification accuracy of {\bf Proposed (BSEC)} with the channel-adaptive modulation is consistently close to the best classification accuracy achieved by the 4-QAM case (see Fig.~\ref{fig:SE}(a)), while improving the spectral efficiency as the SNR increases (see Fig.~\ref{fig:SE}(b)). 
    For example, when ${\sf SNR} = 1$ dB, {\bf Proposed (BSEC)}  with the adaptive modulation provides a two-times higher spectral efficiency than the 4-QAM case, while achieving almost the same classification accuracy.   
    Therefore, our results validate the effectiveness of our channel-adaptive modulation technique in reducing the communication latency while maintaining task performance at the receiver.

    \begin{table}[t]
        \renewcommand{\arraystretch}{1.2}
        \caption{Comparison of the average classification accuracy and spectral efficiency of the proposed JSCC approach for image classification tasks on the MNIST, Fashion-MNIST, and CIFAR-10 datasets.}\label{table:bit_flip_distri}
        \setlength{\tabcolsep}{3pt}
        \footnotesize 
        \centering
        {\begin{tabular}{|cc|cc|cc|cc|cc|} \hline
            \multicolumn{2}{|c|}{Modulation} & \multicolumn{2}{c|}{Adaptive} & \multicolumn{2}{c|}{4-QAM} & \multicolumn{2}{c|}{16-QAM} & \multicolumn{2}{c|}{64-QAM}  \\ \hline \hline

            \multirow{3}{*}{MNIST}  & \multicolumn{1}{|c|}{Acc} & \multicolumn{2}{c|}{$97.18$}  & \multicolumn{2}{c|}{$97.49$}  & \multicolumn{2}{c|}{$92.72$}  & \multicolumn{2}{c|}{$85.69$} \\  \cline{2-10} 
            & \multicolumn{1}{|c|}{SE} & \multicolumn{2}{c|}{$3.79$} & \multicolumn{2}{c|}{$2$} & \multicolumn{2}{c|}{$4$} & \multicolumn{2}{c|}{$6$}   \\  \cline{2-10}
            & \multicolumn{1}{|c|}{CC} & \multicolumn{8}{c|}{$1.57$} \\ \hline \hline

            \multirow{3}{*}{Fashion-MNIST}  & \multicolumn{1}{|c|}{Acc} & \multicolumn{2}{c|}{$86.43$} &  \multicolumn{2}{c|}{$86.94$} &  \multicolumn{2}{c|}{$82.32$} &  \multicolumn{2}{c|}{$75.70$} \\  \cline{2-10} 
            & \multicolumn{1}{|c|}{SE} &  \multicolumn{2}{c|}{$3.79$} & \multicolumn{2}{c|}{$2$} & \multicolumn{2}{c|}{$4$} & \multicolumn{2}{c|}{$6$}   \\  \cline{2-10}
            & \multicolumn{1}{|c|}{CC} & \multicolumn{8}{c|}{$1.57$} \\ \hline \hline

            \multirow{3}{*}{CIFAR-10}  & \multicolumn{1}{|c|}{Acc} & \multicolumn{2}{c|}{$79.55$} &  \multicolumn{2}{c|}{$79.84$} &  \multicolumn{2}{c|}{$77.33$} &  \multicolumn{2}{c|}{$72.83$} \\  \cline{2-10} 
            & \multicolumn{1}{|c|}{SE} &  \multicolumn{2}{c|}{$3.82$} & \multicolumn{2}{c|}{$2$} & \multicolumn{2}{c|}{$4$} & \multicolumn{2}{c|}{$6$}   \\  \cline{2-10}
            & \multicolumn{1}{|c|}{CC} & \multicolumn{8}{c|}{$1.57$} \\ \hline

        \end{tabular}} 
        \label{table:2}
    \end{table}

    Table~\ref{table:2} shows the average classification accuracy and spectral efficiency of the proposed JSCC approach for image classification tasks on the MNIST, Fashion-MNIST, and CIFAR-10 datasets. For CIFAR-10, we set $N$ to be 396. To account for the variability introduced by channel randomness, the performance metrics are averaged over multiple random realizations of the channel coefficients drawn from a uniform distribution, given by $|h|=\sqrt {\sf SNR} \sim {\rm Uniform[0.37,2.5]}$ with $\sigma^2=1$. Each coefficient is assumed to remain constant only during the transmission of 10 images.
    If $\sqrt{\sf SNR} \sim {\rm Uniform}[\gamma_1, \gamma_2]$, the channel capacity is computed according to the Shannon-Hartley theorem \cite{TSC_Cap, TSC_Cap_2}:
    \begin{align}\label{Cap_channel_2} 
        C = \frac{{\rm ln}\frac{(\gamma_2^2+1)^{\gamma_2}}{(\gamma_1^2+1)^{\gamma_1}}+2(\arctan(\gamma_2)-\arctan(\gamma_1))-2(\gamma_2-\gamma_1)}{{\rm ln}2(\gamma_2-\gamma_1)}.
    \end{align}
    Utilizing this fact, in Table~\ref{table:2}, we also provide channel capacity (CC) computed using \eqref{Cap_channel_2} as a performance baseline. 
    

    Table~\ref{table:2} shows that the average classification accuracy achieved with our channel-adaptive modulation technique is almost the same as that achieved with 4-QAM, while providing the average spectral efficiency close to 16-QAM for all the datasets. These results clearly demonstrate the advantages of our modulation technique over a fixed modulation for providing a good latency-performance trade-off for digital semantic communications.
    Table~\ref{table:2} also shows that the latent variables can be transmitted at a rate $2.41$ times faster than the CC for both MNIST and Fashion-MNIST datasets and $2.44$ times faster than the CC for the CIFAR-10 dataset. These results imply that from a semantic perspective, the proposed JSCC approach can maintain a sufficient task performance even if its transmission rate is beyond the theoretical capacity bound determined by the Shannon theorem.

   \section{Conclusion}
   In this paper, we have proposed a novel JSCC approach for enabling channel-adaptive digital semantic communications. To this end, we have first developed a robust demodulation method to prevent frequent bit-flip errors of the binary latent variables while enhancing the expressiveness of a demodulation output. We have then developed a robust training strategy which not only facilitates end-to-end training of the JSCC encoder and decoder but also enhances their robustness and flexibility against diverse channel conditions and modulation orders. We have also devised a channel-adaptive modulation technique that can reduce the communication latency for transmission while maintaining task performance. Using simulations, we have demonstrated that the proposed approach outperforms the existing JSCC approaches in terms of communication latency and task performance. 

   An important direction of future research is to further optimize the performance of the proposed approach through the development of advanced designs for modulation schemes, quantization methods, and loss functions. 
   Another promising research direction is to extend the proposed JSCC approach to accommodate soft-output demodulation for JSCC-based semantic communications. Investigating an extension of the proposed approach to support multi-task multi-user semantic communications, where multiple devices engage in different tasks concurrently, would also be an important direction for future research.

\bibliographystyle{IEEEtran}
\bibliography{Reference}

\end{document}